\documentclass{aastex62}
\usepackage{graphicx}
\usepackage{mathtools}

\received{--}
\revised{-- }
\accepted{ --}
\submitjournal{ApJ}

\shorttitle{Variation of non-thermal line widths}
\shortauthors{Pant et al.}

\begin{document}



\title{\bf{\large{Investigating `dark' energy in the solar corona using forward modeling of MHD waves}}}

\correspondingauthor{Vaibhav Pant}
\email{vaibhav.pant@kuleuven.be, vaibhavpant55@gmail.com}

\author{Vaibhav Pant}
\affiliation{Centre for mathematical Plasma Astrophysics, Department of Mathematics, KU Leuven, Celestijnenlaan 200B, 3001, Leuven, Belgium}

\author{Norbert Magyar}
\affiliation{Centre for mathematical Plasma Astrophysics, Department of Mathematics, KU Leuven, Celestijnenlaan 200B, 3001, Leuven, Belgium}

\author{Tom Van Doorsselaere}
\affiliation{Centre for mathematical Plasma Astrophysics, Department of Mathematics, KU Leuven, Celestijnenlaan 200B, 3001, Leuven, Belgium}

\author{Richard J. Morton}
\affiliation{Department of Mathematics, Physics \& Electrical Engineering, Northumbria University, Newcastle Upon Tyne, NE1 8ST, UK}


\begin{abstract}
It is now well established that the Alfv\'enic waves are ubiquitous in the solar corona. However, the  Alfv\'enic wave energy estimated from the Doppler velocity measurements in the corona was found to be four orders of magnitude less than that estimated from non-thermal line widths. \citet{2012ApJ...761..138M} suggested that this discrepancy in energy might be due to the line-of-sight (LOS) superposition of the several oscillating structures, which can lead to an underestimation of the Alfv\'enic wave amplitudes and energies. \citet{2012ApJ...761..138M} termed this coronal `dark' or `hidden' energy. However, their simulations required the use of an additional, unknown source of Alfv\'enic wave energy to provide agreement with measurements of the coronal non-thermal line widths.
In this study, we investigate the requirement of this unknown source of additional `dark' energy in the solar corona using gravitationally stratified 3D magnetohydrodynamic (MHD) simulations of propagating waves. We excite the transverse MHD waves and generate synthetic observations for the Fe XIII emission line. We establish that the LOS superposition greatly reduces the Doppler velocity amplitudes and increases the non-thermal line widths. Importantly, our model generates the observed wedge-shaped correlation between Doppler velocities and non-thermal line widths. We find that the observed wave energy is only 0.2-1\% of the true wave energy which explains 2-3 orders of magnitude of the energy discrepancy. We conclusively establish that the true wave energies are hidden in the non-thermal line widths. Hence, our results rule out the requirement for an additional `dark' energy in the solar corona.

\end{abstract}


\section{Introduction}
The solar corona is heated to millions of Kelvin, with the mechanism responsible for this heating having evaded researchers for decades \citep{2003A&ARv..12....1W,2006SoPh..234...41K, 2012RSPTA.370.3217P, 2018arXiv181100461C}. One of the possible mechanisms of heating is the dissipation of magnetohydrodynamic (MHD) waves in the solar atmosphere \citep{ 2006SoPh..234...41K,2015RSPTA.37340261A}. MHD waves and their different wave modes have been ubiquitously observed in the solar atmosphere by both space and ground-based instruments \citep{2007SoPh..246....3B}. It was suggested that hydromagnetic waves in the solar atmosphere can produce non-thermal broadening of the emission lines \citep{1973ApJ...181..547H,2008ApJ...676L..73V}. The evidence of the broadening of transition region emission lines in the quiet sun and coronal holes was first provided by the spectrograph on the {\it Skylab} \citep{1976ApJ...205L.177D, 1976ApJS...31..417D, 1976ApJS...31..445F}.
Early observations of non-thermal broadening of coronal emission lines (formed at temperatures $\sim$~10$^{6}$~K) by \citet{1990ApJ...348L..77H} pointed to the possible existence of the Alfv\'en waves in the solar atmosphere. Later, \citet{1998A&A...339..208B} and \citet{1998SoPh..181...91D} used the Solar Ultraviolet Measurements of Emitted Radiation (SUMER) spectrometer on-board {\it Solar and Heliospheric Observatory} (SOHO) to study the variation of line widths of the Si VIII emission line at different locations in northern and southern polar coronal holes. These authors observed that the line width of the Si VIII emission lines increased from 27 to 46~km~s$^{-1}$ when the distance above the solar limb increased to 17~Mm to 175~Mm, respectively. They computed the energy flux in the Alfv\'en waves and found that it is slightly less than the energy flux (8$\times$10$^{5}$~ergs~cm$^{-2}$ s$^{-1}$) required to balance total coronal energy losses in coronal holes \citep{1977ARA&A..15..363W}. Similar studies on the nature of non-thermal broadening of coronal emission lines were carried out using the Coronal Diagnostic Spectrometer (CDS) \citep{2005A&A...436L..35O}, Ultraviolet Coronagraph Spectrometer (UVCS) on-board SOHO \citep{1997ApJ...476L..51O,1999ApJ...510L..59K,2016SSRv..201...55A} and the Extreme Ultraviolet Imaging Spectrometer (EIS) on-board {\it Hinode} \citep{2007ApJ...667L.109D,2009A&A...501L..15B,2012ApJ...753...36H}. The estimated energy flux was found to be just enough to balance the energy losses in the polar open-field regions of the solar corona. Additionally, several 1D \citep{1996ApJ...465..451L, 2003ApJ...596..646O, 2006JGRA..111.6101S,2007ApJS..171..520C,2013ApJ...778..176O,2014ApJ...782...81V,2017ApJ...845...98O}, 2.5D \citep{1997ApJ...476..357O, 1998JGR...10323677O}, and 3D \citep[using reduced MHD;][]{2011ApJ...736....3V,2017ApJ...849...46V} wave-based models driven purely by Alfv\'en waves have been somewhat successful in explaining the large non-thermal widths of coronal emission lines and acceleration of fast solar winds in open and closed magnetic field regions.

\bigskip
Resolved measurements of the propagating Alfv\'enic waves came from observations of the chromosphere and transition region using imaging data from the Solar Optical Telescope (SOT) on board {\it Hinode} \citep{2007Sci...318.1574D} and the Atmospheric Imaging Assembly (AIA) on-board the {\it Solar Dynamics Observatory} (SDO) \citep{2011Natur.475..477M}. These authors reported waves with amplitudes of $\sim$~20~km~s$^{-1}$, suggesting that they are capable of providing the energy flux of 100-200~W~m$^{-2}$ to accelerate fast solar wind and balance radiative losses in quiet corona. The coronal counterpart to these propagating Alfv\'enic waves was observed using Doppler velocity data \citep{2007Sci...317.1192T, 2008SoPh..247..411T, 2009ApJ...697.1384T, 2015NatCo...6E7813M} from the Coronal Multi-Channel Polarimeter (CoMP \citealp{2008SoPh..247..411T}), and through direct measurements with SDO/AIA \citep{2014ApJ...790L...2T, WEBetal2018,2019NatAs...3..223M}. Both decaying \citep{1999Sci...285..862N,2002SoPh..206...99A} and decayless \citep{2013A&A...552A..57N,2013A&A...560A.107A} transverse oscillations have been observed in the solar atmosphere through direct imaging. The CoMP data can also measure coronal line widths, providing estimates of the non-thermal component that is comparable to previously reported values. Surprisingly, the measured Doppler velocity fluctuations only have amplitudes of $\sim$0.5~km~s$^{-1}$, which suggests the Alfv\'enic waves only have an energy flux of $\sim$0.01~W~m$^{-2}$. Furthermore, in a comparative study, \citet{2015NatCo...6E7813M} reported an average velocity amplitude of $\sim$14~km~s$^{-1}$ measured with SDO at $1.01~R_\odot$, which is at a lower height in the corona than where CoMP Doppler velocities are measured. Thus, there seems to be an apparent discrepancy in the wave energy estimated using the CoMP Doppler velocities compared to those estimated using non-thermal line widths and the imaging measurements from SOT and AIA.

This discrepancy was investigated by \cite{2012ApJ...761..138M}, who used a Monte Carlo method to forward model the emission spectrum generated by several oscillating structures (termed as `threads') along the line-of-sight (LOS). We note that the simulations were effectively a toy model of how Alfv\'enic waves would impact spectral lines and no MHD simulations were used. It was suggested that small Doppler velocities and large non-thermal line widths are the consequences of unresolved swaying motions of threads along the LOS due to optically thin solar corona. In addition to this, the authors used CoMP to demonstrate a correlation between root mean square (rms) Doppler velocities and mean non-thermal line widths that appeared as a wedge-shape. The authors could explain the wedge-shaped correlation using their model, but, only by including an additional component of non-thermal broadening whose origin was not known. Finally, these authors suggested that the `dark' or `hidden' energy, which is not observed by direct imaging, could be residing in the non-thermal line widths. The effects of superposition of coronal loops along the LOS on the wave amplitudes was also investigated using a 3D MHD model \citep{2012ApJ...746...31D}. In this study, the authors found that the kinetic energy measured from the LOS Doppler velocities is an underestimate of the total kinetic energy present in the model.

In spite of the developments outlined above, little work has been done to investigate the LOS effects on Doppler velocities and line widths in the solar corona in the context of the CoMP. This displays the need for an in-depth study of the wave propagation in the solar corona using MHD models, examing whether they can generate the observed values of non-thermal widths of emission lines and their variation with height through the corona. Additionally, the model should also be able to produce a wedge-shaped correlation between Doppler velocities and non-thermal line widths.

\medskip
In this work, we investigate the correlation between rms Doppler velocities and non-thermal line widths in open magnetic field regions using 3D MHD simulations of propagating waves. Further, we explore the requirement of an additional non-thermal broadening, which is needed to explain the wedge-shape correlation in  \cite{2012ApJ...761..138M}. We forward model our MHD simulations for the Fe XIII emission line and examine the variation of the non-thermal line widths with height in the solar atmosphere. The solar plasma is inhomogeneous and gravitationally stratified, which leads to the reflection \citep{1958ApJ...127..459F,1978SoPh...56..305H} and non-linear interaction of waves \citep{1999ApJ...523L..93M,2007ApJS..171..520C} propagating through it. Recently, \citet{2017NatSR...714820M} have studied the effects of perpendicular inhomogeneities on unidirectionally propagating MHD waves and reported that such inhomogeneities lead to a generalised phase mixing that generates a turbulence-like behaviour. These authors termed this `uniturbulence' since it is produced by unidirectionally propagating Alfv\'enic waves. \citet{2019arXiv190102676K} have investigated the effects of gravitational stratification on the heating of coronal loops using 3D MHD simulations. These authors have reported that the inclusion of gravity increased the average temperature near the footpoint and the apex of the coronal loop compared to the simulations where gravity was excluded \citep[see also][]{2017A&A...604A.130K}. Since the gravity and plasma inhomogeneities alter the wave propagation, it is crucial to study their effects on the observed properties of waves. Thus we extend the MHD model of unidirectional propagating waves in a perpendicularly inhomogeneous plasma \citep{2017NatSR...714820M, 2018ApJ...856..144M} by including gravity and investigate the effects LOS superposition of waves on the observables such as Doppler velocities and line widths. The paper is structured as follows. The observations using CoMP are reported in section~\ref{obs}. The choice of parameters (in the model) and numerical set-up is discussed in Section~\ref{sec2}. Forward modeling using Fe XIII emission line is described in Section~\ref{fm}. Section~\ref{sec4} outlines the analysis and results which is followed by a discussion and conclusions in Section~\ref{sec5}.

\section {Observations}
\label{obs}
Here we use observations taken with the CoMP instrument on the 27 March 2012. The data set has been used previously in \citet{2015NatCo...6E7813M, 2016ApJ...828...89M, 2019NatAs...3..223M}, where discussion of any additional post-processing to the Level 2 data files are given in detail. CoMP measures intensity at three positions across the Fe XIII 10747~{\AA} line, from which estimates for the Doppler shifts and line widths can be made. The rms Doppler velocities and non-thermal line widths are calculated from these quantities, where we follow the prescription in \citet{2012ApJ...761..138M} to calculate the non-thermal component of the line widths (and permit direct comparison between results).

\begin{figure}[ht!]
\centering
\includegraphics[scale=0.7]{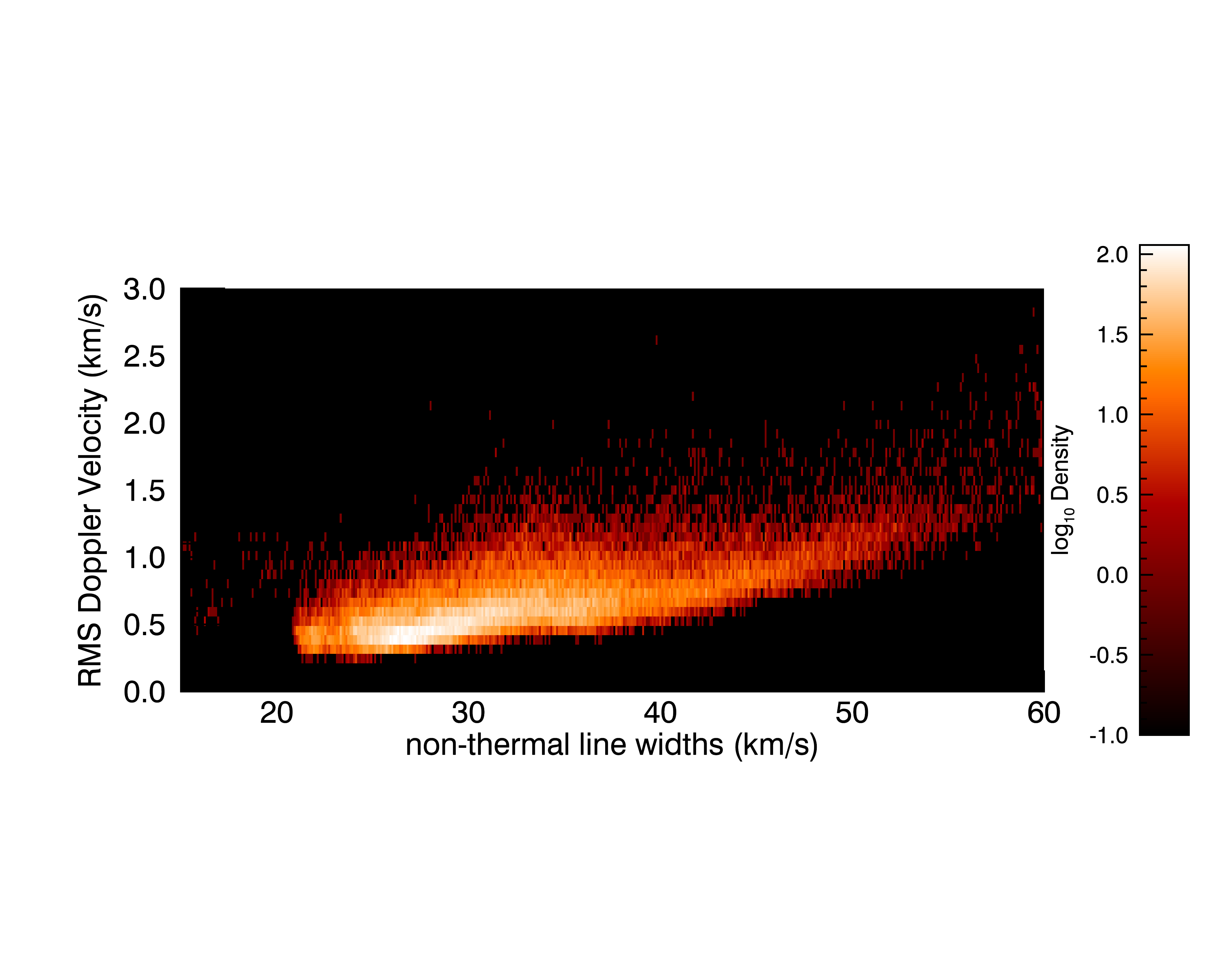}\\
\includegraphics[scale=0.6]{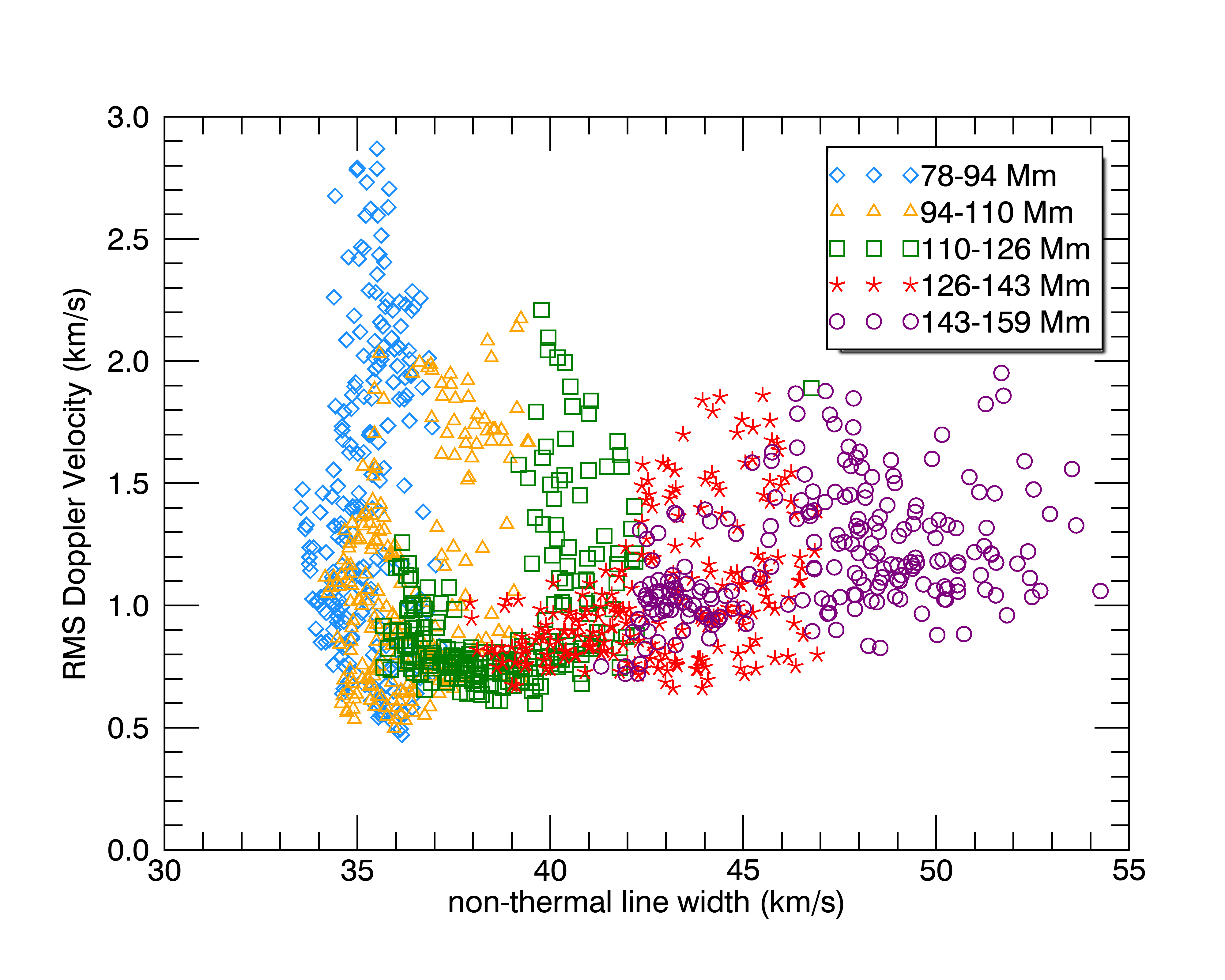}
\caption{{\it Top}: Two dimensional histogram displaying the joint distribution of the rms Doppler velocities and mean non-thermal line widths taken from data covering the entire CoMP FOV. The colorbar indicates the density of the points. {\it Bottom}: Scatter plot showing the variation of the rms Doppler velocities with mean non-thermal line widths in a coronal hole where magnetic fields are aligned radially outwards. The different colors correspond to the different height ranges above the limb.}
\label{fig0}
\end{figure}

In Figure~\ref{fig0}, we plot a two-dimensional histogram of the non-thermal line width against rms Doppler velocity for the entire field-of-view (FOV). It is seen that a similar wedge shape is obtained to that in \citet{2012ApJ...761..138M}. There are some differences between the location of the wedge and its exact shape. However, this variation can likely be attributed to differences in the coronal magnetic field and plasma conditions, which leads to different wave properties \citep[e.g,][]{2016ApJ...828...89M, 2019NatAs...3..223M}.

\medskip
It is well known that the density of the corona decreases as a function of height. CoMP data provides measurements between $1.05~R_\odot$ to $1.3~R_\odot$, which corresponds to approximately 2 pressure scale-heights for a 1.6~MK plasma. Furthermore, it is known that the amplitude of Alfv\'enic modes depends on the density, which from WKB theory goes as $v\propto\langle\rho\rangle^{-1/4}$. Hence, we should expect that Doppler velocities and non-thermal line widths measured with CoMP show some dependence with height. An increase in amplitude with height is visible in the full FOV images of Figure~1 in \citet{2012ApJ...761..138M}. Moreover, previous observations of coronal holes, with CoMP and other spectrometers, have demonstrated an increase in amplitude for both rms velocity and non-thermal line widths (for  altitudes $< 1.2R_\odot$) that are broadly in agreement with the WKB theory \citep[e.g,][]{2012ApJ...753...36H,2016ApJ...828...89M}. However, we note, that the observed increase in amplitude in other regions of the corona (i.e. away from coronal holes) will not match the expected amplification for WKB Alfv\'enic waves due to density stratification, as the waves are known to undergo some form of damping in the corona (e.g. \citealp{2010ApJ...718L.102V}, \citealp{Tiwari2019}). It is worth highlighting that kink waves in compressible plasmas have mixed properties that are similar to the surface Alfv\'en waves \citep{2009A&A...503..213G, 2012ApJ...753..111G}, thus several authors used the term `Alfv\'enic' to describe such waves. In general such waves are transverse and largely incompressible. In this work, we call these waves as transverse MHD waves.

To examine how the variation in wave amplitude with height influences the wedge shape, we show a portion of the data from a coronal hole region in the lower panel of Figure~\ref{fig0}. In this region the magnetic field is almost radial, and hence enables us to unambiguously show the influence of change in density with height on both the rms velocity and non-thermal line widths. Figure~\ref{fig0} reveals that part of the contribution to the wedge shape, affecting both the range of non-thermal width values and the positive correlation with rms Doppler velocity, is due to the dependence of the transverse MHD wave amplitude on height\footnote{The larger values of non-thermal line widths at 78-94~Mm are likely due to the influence of scattered light from the CoMP occulter.}. We note that this consideration was neglected from the analysis of \citet{2012ApJ...761..138M} and, as such, raises questions about the inferred values of wave properties from their Monte Carlo simulations.

\section {Numerical setup}
\label{sec2}
To understand the transverse MHD wave propagation in open-field regions, we perform an ideal 3D MHD simulation using MPI-AMRVAC that solves hyperbolic partial differential equations in near conservative form \citep{2014ApJS..214....4P}. The following equations are solved in cartesian geometry for a grid size of 128$\times$512$\times$512 that span 50~Mm$\times$5~Mm$\times$5~Mm (see Figure~\ref{fig1}~a).\\
\begin{equation}
\begin{aligned}
\frac{\partial \rho}{\partial t} + {\bf \nabla} . (\rho {\bf v})=0, \\
\frac{\partial ({\rho {\bf v}})}{\partial t} + \nabla.(\rho {\bf v v} - {\bf BB}/\mu_{0}) + \nabla (p + {\bf B}^{2}/2\mu_{0}) - \rho {\bf g} = 0,\\
\frac{\partial E}{\partial t} + \nabla.({\bf v}E - {\bf BB.v}/\mu_{0} + {\bf v}.(p + {\bf B}^{2}/2\mu_{0})) - \rho{\bf v.g}= 0,\\
\frac{\partial B}{\partial t} - \nabla \times ({\bf v} \times {\bf B}) = 0,\\
\nabla.{\bf B} = 0,
\end{aligned}
\end{equation}
where $\bf{B}$ is the magnetic field, $\bf{v}$ is the plasma velocity, $\rho$ is the density, $p=\rho\frac{k_{b}}{\mu m_{H}}T$, is the gas pressure and $E$ is total energy density defined as, $E=\frac{p}{\gamma -1} + \frac{\rho {\bf v}^{2}}{2} + \frac{{\bf B}^{2}}{2\mu_{0}}$. Furthermore, $\mu_{0}$ is the magnetic permeability in the free space, {\bf g} is the acceleration due to gravity pointing along the negative $x$ axis, $\mu$ is taken to be 0.6 for coronal abundance, $m_{H}$ is the mass of the proton, $k_{b}$ is the Boltzmann constant, and $\gamma$ is chosen to be 5/3.

\subsection{Initial Conditions}
 A uniform grid without any mesh refinement is employed for performing the simulations. The spatial resolution along the $x$, $y$, and $z$ axis is 0.39 Mm, 0.01 Mm, and 0.01 Mm, respectively, and the $x$ axis defines the vertical direction. Initially, the simulations are set up assuming a vertically isothermal atmosphere, meaning the temperature is constant with height. Thus, $\rho$ is an exponentially decaying function of the height due to the gravitational stratification. Further, in the initial setup, inhomogeneities in density are randomly placed transverse ($y-z$ plane) to the direction of the magnetic field according to the following equation,
\begin{equation}
\rho(x,y,z)= \left(\rho_{0}+\sum_{i=0}^{50} A_{i}\exp^{-[(y-y_{i})^{2} + (z-z_{i})^{2} ]/2\sigma_{i}^{2}}\right)\exp^{-x/H(y,z)}.
\label{eq2}
\end{equation} 
$H (y,z)$ is the scale height that depends on the temperature which is different at different locations in the $y-z$ plane. We choose $\rho_{0}=2\times$10$^{-13}$~kg~m$^{-3}$. The magnitude of the inhomogeneity is given by $A_{i}$, which is randomly drawn from a uniform distribution of [0, 5]~$\rho_{0}$. The spatial extent of the inhomogeneity is controlled by $\sigma_{i}$, which is randomly chosen from [0, 250]~km. The spatial location of the inhomogeneity $y_{i}$ and $z_{i}$ are chosen within the simulation domain {\it i.e}, drawn randomly from a uniform distribution of [-2.5, 2.5]. We set the gas pressure in the $y-z$ plane to constant, hence Eq.~\ref{eq2} also indirectly determines the initial temperature structure. The initial magnetic field is assumed to be uniform and vertical, $\textbf{B}=B\bf{\hat{x}}$, and its strength is prescribed to be 5~G. This value of the magnetic field strength is typical for coronal holes \citep{1990CoPhR..12..205H}. We assume a low beta ($\beta$) plasma with $\beta=0.15$.

Longitudinal and transverse sections of the initial setup are shown in panels (b) and (c) of Figure~\ref{fig1}. Figure~\ref{fig2} shows the variation of the density with height in the initial setup at the location ($y=0$, $z=-0.9$) marked with the dashed line in Figure~\ref{fig1} (b). Corresponding to the location of the dashed line in Figure~\ref{fig1}, the scale height is 40 Mm (see, Figure~\ref{fig2}).

\subsection{Initial evolution}
Since the initial setup described above is not in the pressure equilibrium, we evolve the simulations initial setup for $\sim$100~s before implementing any driving. We use a TVD second-order solver and a Woodward slope limiter to solve for $p$, ${\bf v}$, $\rho$, and ${\bf B}$. Moreover, Powell's scheme is employed to ensure a divergence-free magnetic field. For this stage we employ open boundary conditions across all boundaries, such that any MHD waves generated can leave the simulation domain. This evolutionary step allows the initial state of the system to relax to a state of pressure equilibrium. As the plasma relaxes, the density inhomogeneities and the magnetic field expand in response (see, Figure~\ref{fig2}). The magnetic field becomes vertically stratified and concentrates at the density inhomogeneities due to the lower gas pressure in these regions. From here on, we use $t=0$ to denote the time when the simulations have reached at the pressure equilibrium. An example of the temperature and density structure in the $y-z$ plane is shown in the top panel of Figure~\ref{emission}.


\subsection{Wave excitation}
Once the simulation reach pressure equilibrium, we excite transverse MHD waves by driving the entire bottom boundary ($x$=0~Mm) perpendicular to the direction of the background magnetic field ({\it i.e,} in $y-z$ plane). In contrast to the initial evolution, boundaries in the $y$ and $z$ directions are now set to be periodic. However, the top boundary of the simulation domain is kept open, so that transverse MHD waves can leave the domain.

\begin{figure}[ht!]
\includegraphics[scale=0.5]{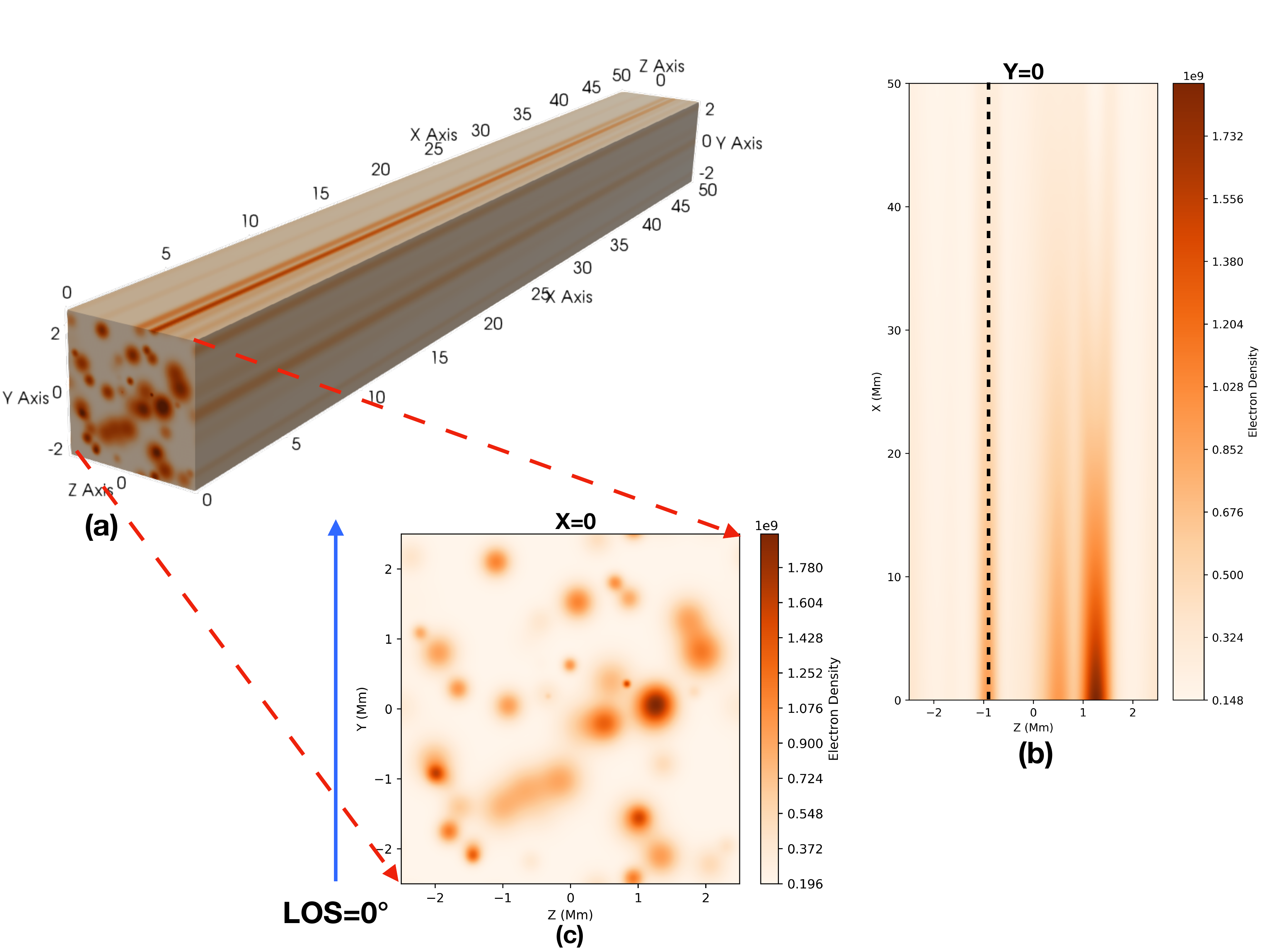}
\caption{(a) Full simulation cube at $t$=0. The initial bottom boundary is shown in panel (c). (b) A longitudinal cut at $y$=0 Mm. Density stratification is evident along the $x$ axis. The black dashed line represents the region at $y$=0 and $z$=-0.9 Mm which is used for further analysis. (c) A transverse cut ($y-z$ plane) at $x$= 0 Mm. The arrow in blue indicates the direction of integration when the LOS is chosen to be 0$^{\circ}$ for performing forward modeling with FoMo. Electron density is color coded in the units of cm$^{-3}$. }
\label{fig1}
\end{figure}

\begin{figure}[ht!]
\centering
\includegraphics[scale=0.55]{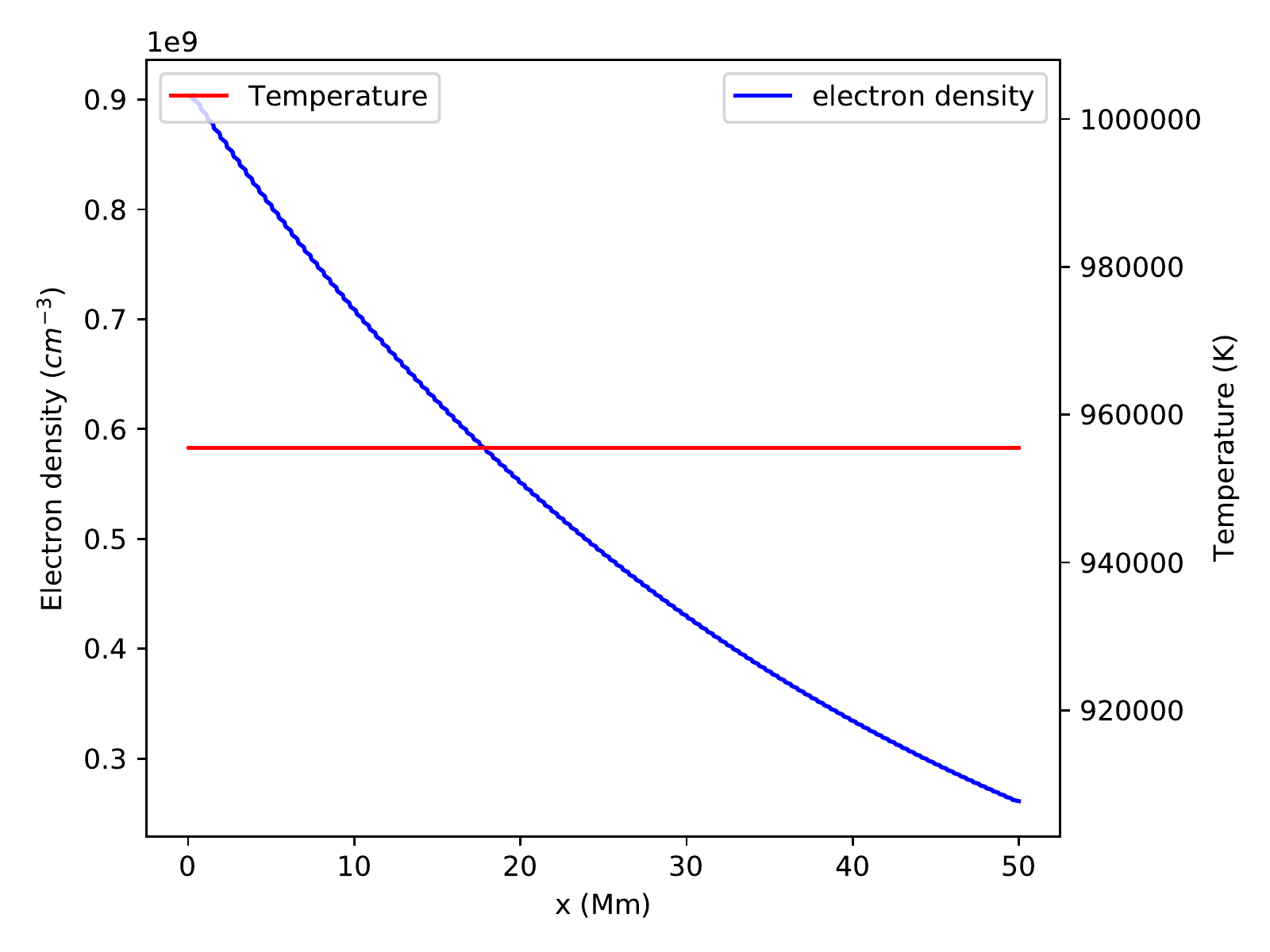}
\includegraphics[scale=0.55]{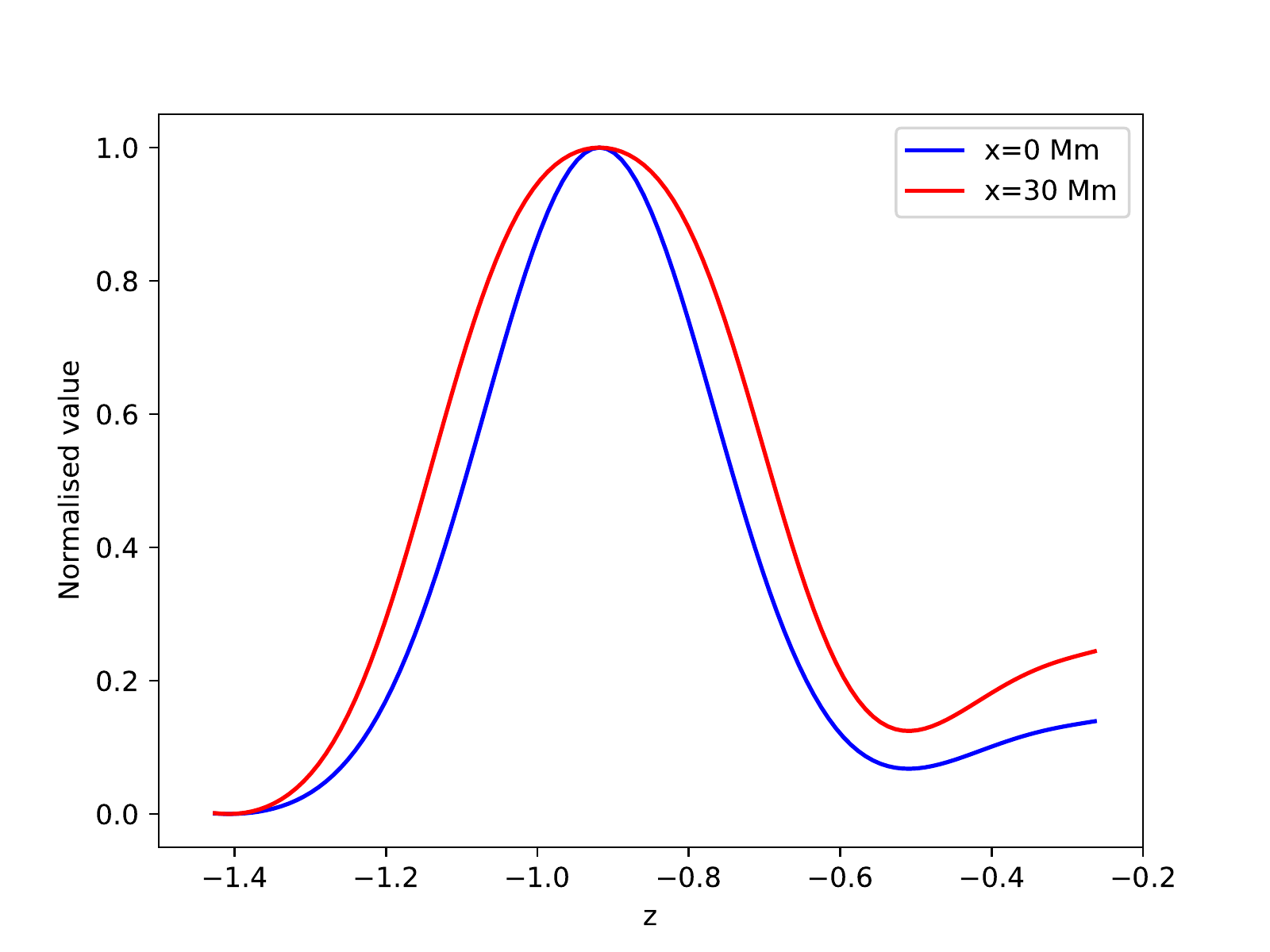}
\caption{{\it Left}: Variation of density and temperature with height ($x$ axis) for the initial set up at the location marked with the dashed curve in Figure~\ref{fig1} (b). {\it Right}: Expansion of density inhomogeneities at different heights.}
\label{fig2}
\end{figure}

The bottom boundary is driven uniformly by a velocity driver composed of a superposition of ten velocity drivers with different periodicities. The $y$ and $z$ components of the velocity driver are given by the following equations,
\begin{equation}
\begin{aligned}
v_{y}(x=0,t)=\sum_{i=1}^{10} U_{i} \sin(\omega_{i} t),\\
v_{z}(x=0,t)=\sum_{i=1}^{10} V_{i} \sin(\omega_{i} t),\\
\end{aligned}
\end{equation}
where periods (hence, $\omega$'s) are chosen from the observed log-normal distribution \citep{2014ApJ...790L...2T, WEBetal2018,2015NatCo...6E7813M,2019NatAs...3..223M}. The magnitude of $U_{i}$ and $V_{i}$ are randomly chosen from a uniform distribution of $[-U_{o},U_{o}]$. In this work, we perform simulations for three different values of $U_{o}$: $\frac{5}{\sqrt{2}}$~km~s$^{-1}$, $\frac{11}{\sqrt{2}}$~km~s$^{-1}$, and $\frac{22}{\sqrt{2}}$~km~s$^{-1}$. The root mean square (rms) value of velocity driver ($v_{rms}$) averaged over the entire bottom boundary corresponding to above three cases are $\sim$7~km~s$^{-1}$, $\sim$15~km~s$^{-1}$, and $\sim$26~km~s$^{-1}$, respectively. The $v_{rms}$ averaged over the entire bottom boundary ($x=0, y, z$) is computed using the following relation. 
\begin{equation}
\begin{aligned}
v_{rms}=\left \langle\sqrt{\frac{\sum_{t=0}^{T-1} [ v_{y}^{2}(x=0,y,z,t) + v_{z}^{2}(x=0,y,z,t) ]}{T} } \right \rangle_{y,z}; T=50.
\label{eq4}
\end{aligned}
\end{equation}
Here $T$ is the total duration of the simulations. All wave simulations are performed for 1000 s with a cadence of 20 s, thus giving 50 snapshots for each run. Angular brackets represents the average over $y-z$ plane. Henceforth, throughout the manuscript, $v_{rms}$ represents the rms velocity of the driver at the bottom boundary. Due to the difference in periodicity and amplitudes, the resultant velocity field has a varying phase with time. An animation corresponding to the panel (c) of Figure~\ref{fig1} is available for the case where the lower boundary of the simulation domain is driven by a velocity driver with $U_{o}$=$\frac{11}{\sqrt{2}}$~km~s$^{-1}$. The average sound speed ($c_{s}$) and Alfv\'en speed (v$_{A}$) at $t=0$ are $\sim$~120~km~s$^{-1}$ and 500~km~s$^{-1}$ respectively. 
Since $v_{rms}$ $<$ $c_{s}$ $<$ v$_{A}$ for all cases, the excited waves are in a linear regime and weak compressible limit (or largely incompressible).

\section {Forward modeling with FoMo}
\label{fm}
To compare the results of the simulations with observations taken with the CoMP (Section~\ref{obs}), we need to convert physical variables obtained from the simulations (e.g., density, temperature ($T$), velocity) to spectroscopic observables such as specific intensity (function of wavelength) for the Fe XIII emission line centred at 10749~\AA, from which the Doppler velocities and line widths can be derived. Since the solar corona is optically thin, the specific intensity ($I$) is computed by adding the emission ($\epsilon$) of different structures along the LOS. We use the FoMo tool developed for the forward modeling of the optically thin emission from the coronal plasma \citep{refId0,10.3389/fspas.2016.00004}, specifically FoMo-C (adapted for C++), to compute intensities, $I(\lambda,x,z,t$), and generate synthetic observations comparable to CoMP. The line formation temperature and thermal width of the Fe XIII emission line are $\sim$1.6~MK and $\sim$21.78~km~s$^{-1}$ respectively. For calculating the emission in Fe XIII (10749~\AA), we use the CHIANTI atomic database \citep{1997A&AS..125..149D}, a coronal abundance of the Fe XIII relative to the hydrogen, and assume that the emitting plasma is at ionisation equilibrium. Specifically, the contribution function, $G(n_{e},T)$, for the Fe~XIII is computed using the CHIANTI. Further, the emission is computed at every location using the following relation; 
\begin{equation}
    \epsilon(x,y,z,t)=\frac{A_{b}}{4 \pi}n_{e}^{2}(x,y,z,t)G(n_{e},T),
    \label{emiss}
\end{equation}
where $A_{b}$ is the coronal abundance and $n_{e}$ is the electron density \citep[see][]{refId0,10.3389/fspas.2016.00004}. Computation of $G(n_{e},T)$ requires the knowledge of the rates of electron 
excitation/de-excitation, proton excitation/de-excitation, and photoexcitation/stimulated emission \citep[see][]{1997A&AS..125..149D,2003ApJS..144..135Y}. We note that photoexcitation does not influence our results (see Appendix A).

The upper right panel in Figure~\ref{emission} shows the $\epsilon$ computed using Eq.~\ref{emiss} at $x$=20~Mm and $t$=0. It is evident that the $\epsilon$ is weak in the regions of high density and low temperature ({\it i.e,} inside the density inhomogeneties) or low density and high temperatures ({\it i.e,} outside the density inhomogeneities). However the emission is maximum where temperatures are $\sim$1.6~MK ({\it i.e,} at the boundaries of the inhomogeneities). This happens because the $G(n_{e},T)$ sharply varies with temperature and peaks at $\sim$1.6~MK. It should be noted that after performing the LOS integration of $\epsilon$ in the $y-z$ plane, inhomogenities still appear brighter than the surroundings, even though the emission is less inside the inhomogeneties. The bottom right panel of Figure~\ref{emission} shows the emission at $x$=20~Mm and $t$=45 when the turbulence has developed. This lead to the formation of fine scale structures in the synthetic images obtained after LOS integration of $\epsilon$.

\begin{figure}[ht!]
\includegraphics[scale=0.45]{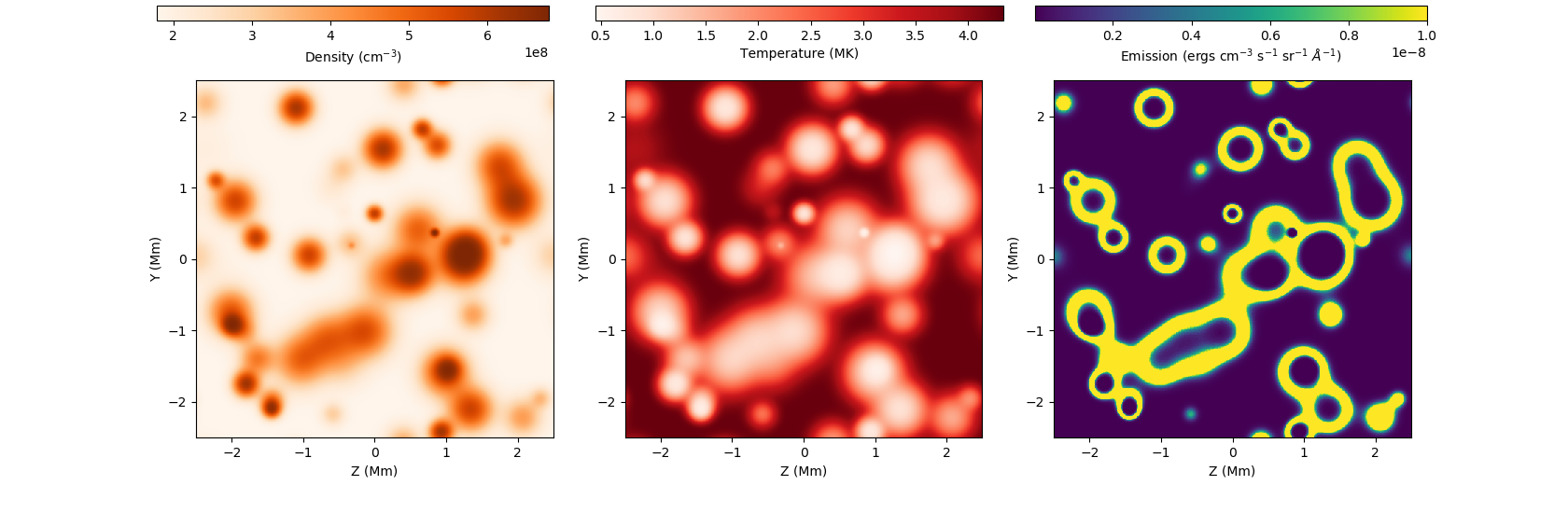}
\includegraphics[scale=0.45]{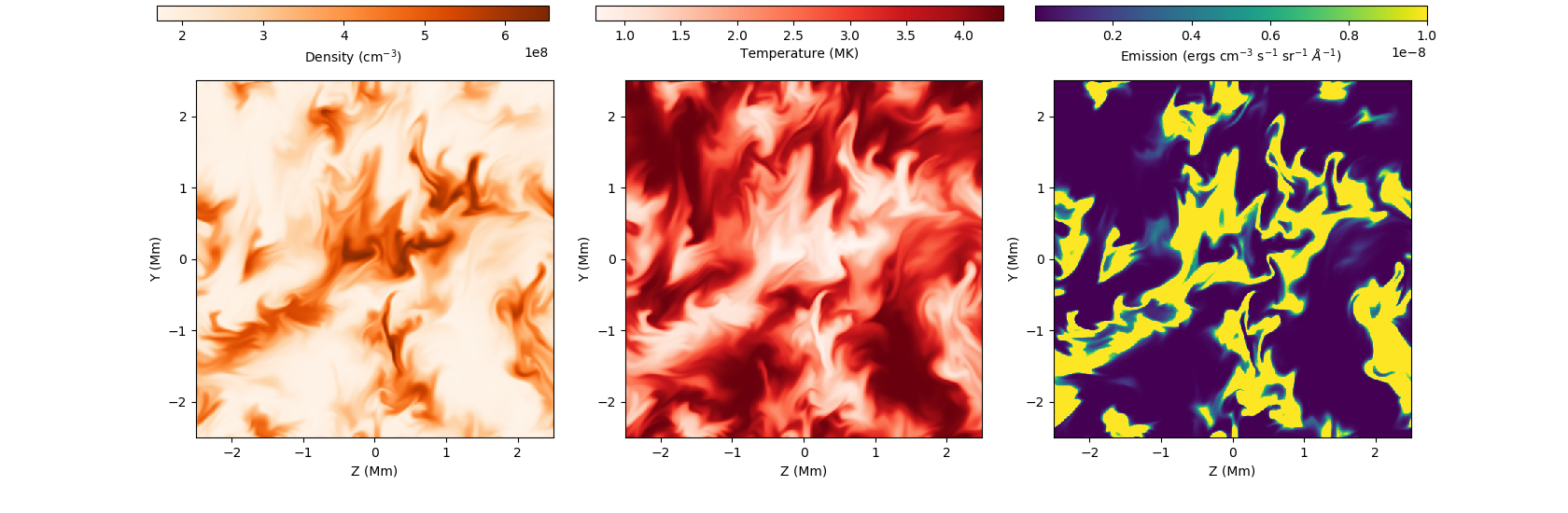}
\caption{{\it Top}: Electron density ($n_{e}$), temperature and emission ($\epsilon$) at $x$=20 Mm and $t$=0 in Fe XIII emission line centred at 10749~\AA~obtained after the application of FoMo on the simulation with $U_{0}=\frac{11}{\sqrt{2}}$~km~s$^{-1}$. {\it Bottom}: Same as top panel but at $x$=20 Mm and $t$=45.}
\label{emission}
\end{figure}

\smallskip

We choose twelve different LOS directions (0$^{\circ}$, 15$^{\circ}$, 30$^{\circ}$, 45$^{\circ}$, 60$^{\circ}$, 75$^{\circ}$, 90$^{\circ}$, 105$^{\circ}$, 120$^{\circ}$, 135$^{\circ}$, 150$^{\circ}$, and 165$^{\circ}$) through the $y-z$ plane perpendicular to the $x$ axis. After the application of the FoMo, LOS integration reduces the data cube to two spatial dimensions and provides the monochromatic specific intensity, $I_\lambda$ from which we calculate the total intensity, $I$, line widths ($\sigma$), and Doppler shifts ($\lambda_{D}$). As an example, the direction of integration for LOS=0$^{\circ}$ is along the $y$ axis (as  shown in Figure~\ref{fig1} (c)) and we obtain $I_\lambda(x,z,t$). Then, at all instances and locations we compute, $I$, $\sigma$, $\lambda_D$ by taking the moments of $I_\lambda(,x,z,t$), as given by the following relations:
\begin{equation}
\begin{aligned}
I(x,z,t)= \int_{\lambda} I_\lambda(x,z,t)d\lambda ,\\
\lambda_{D}(x,z,t)=\frac{\int_{\lambda} \lambda I_\lambda(x,z,t)d\lambda}{ \int_{\lambda} I_\lambda(x,z,t)d\lambda} - \lambda_{0} ,\\
\sigma(x,z,t)=\sqrt{\frac{ \int_{\lambda} (\lambda-\lambda_{D})^{2} I_\lambda(x,z,t)d\lambda}{\int_{\lambda} I_\lambda(x,z,t)d\lambda} }.
\label{eq5}
\end{aligned}
\end{equation}
Here $\lambda_{0}$ is the location of the peak of the Fe XIII emission line which is 10749~\AA. Further, exponential line width ($\sigma_{1/e}=\sqrt{2}\sigma$) is computed and subsequently, converted to the velocity. Similarly the Doppler shifts are also converted to velocities ($v_{D}$). A similar method is adopted for deriving specific intensities, Doppler velocities and exponential line widths for other LOSs.

\begin{figure}[ht!]
\centering
\includegraphics[scale=0.42]{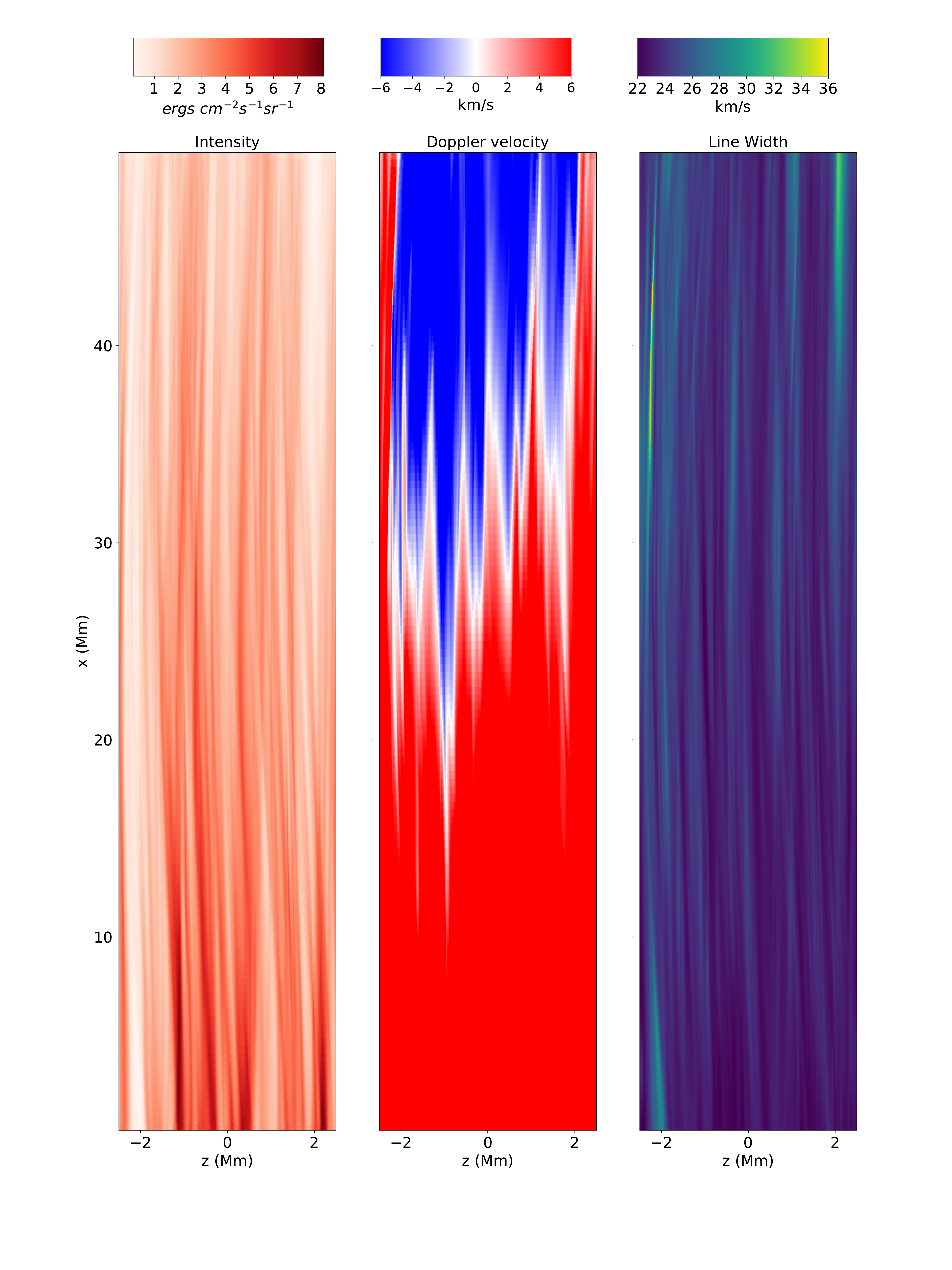}
\caption{Synthetic image of total intensity, Doppler velocity, and total line widths at LOS=0$^{\circ}$ and $t$=27 for Fe XIII emission line centred at 10749~\AA~obtained after the application of FoMo on the simulation with $U_{0}=\frac{11}{\sqrt{2}}$~km~s$^{-1}$.}
\label{fig4}
\end{figure}

The synthetic images of the total intensity, Doppler velocity, and exponential line width for LOS=0$^{\circ}$ and $U_{o}$=$\frac{11}{\sqrt{2}}$~km~s$^{-1}$ are shown in Figure~\ref{fig4}. An animation of this figure is available. While performing the FoMo, we choose a spectral resolution of 4~km~s$^{-1}$ and the spatial resolution is kept as that of the simulation cube. To compare the synthetic observations with the observations, we degrade the spatial resolution of the synthetic observations to that of CoMP ($\sim$3200~km). However, for simplicity, we maintain the same spectral resolution, which is higher than the spectral resolution of the CoMP ($\sim33$~km/s).\footnote{We mention that \citet{2012ApJ...761..138M} did not find any significant differences in the $v_{rms}$ and $\sigma_{1/e}$ when synthesizing data with both low and high spectral resolutions.}

\section {Analysis and Results}
\label{sec4}
In the left panel of Figure~\ref{fig5} we show an example of the Fe XIII emission line profile (blue) at the start of the simulation ($t$=0), when the velocity driver at the bottom boundary is not applied. The width of the initial line profile shown is $\sim$22.5~km~s$^{-1}$, which is slightly greater than the thermal width of the Fe XIII line ($\sim$21.78~km~s$^{-1}$). This difference is due to the temperature and density inhomogeneities along the LOS. Overplotted in green is the line profile averaged over 400 s (20 frames)\footnote{We chose 20 frames for averaging the spectra because the average time period of oscillations in our simulation is 400 s (or 20 frames).}, where there is additional broadening due to the time-averaged behaviour of the transverse wave motions.   


The right panel of Figure~\ref{fig5} presents the line profile at the same location but with a larger amplitude driver ($U_{o}$=$\frac{22}{\sqrt{2}}$~km~s$^{-1}$). The width of the initial line profiles (blue) are the same in both panels, while the time-averaged line profile (green) is significantly broader than for the simulation with the lower amplitude driver. The larger wave amplitude produces a broader emission line profile when integrated over the time period of oscillations, provided the medium is optically thin, because of the superposition of the shifted spectra due to the incoherent and spatially unresolved swaying motions of the structures.

\subsection{Superposition of line profiles}
Since the solar corona is optically thin, the spectrum observed at a given location in the plane-of-sky (POS) is the superposition of the spectra of different structures oscillating with random phases and different polarisations along the LOS. Given the relatively short length of our simulation box, we effectively stack multiple realisations of the simulation along the LOS to mimic the corona. The following procedure is apodted to obtain random segments (from the synthetic observations) that are oscillating in random phases and with random polarisations.

First, we randomly choose several LOSs from the twelve different LOS (as described in Section~\ref{fm}) with a uniform probability of choosing any LOS. This is equivalent to choosing different polarisations of the waves. Next, we randomly chose several start times ($t_{0}$) of oscillations and considered twenty consecutive frames for every start time. The random choices of the start time are equivalent to assuming that different structures are oscillating with different phases. A segment, $I_\lambda(x,z',t_{0,j})$, is then defined as a 2D projection of the simulation cube, obtained by choosing a random start time ($t_{0,j}$) of oscillation and integrated over a random LOS ($LOS_{j}$) direction. $z'$ represents the arbitrary direction that is perpendicular to the chosen random LOS and confined in the $y-z$ plane. For example, if $LOS_{j}$=0$^{\circ}$, $z'_{j}=z$ because the LOS is along $y$ axis. Since we have several LOSs and start times, we get several such random segments. Finally, we superimpose the intensities of all such random segments to obtain a resultant intensity that will be used for further analysis. This method can be understood with the following relation:

\begin{figure}[t!]
\centering
\includegraphics[scale=0.55]{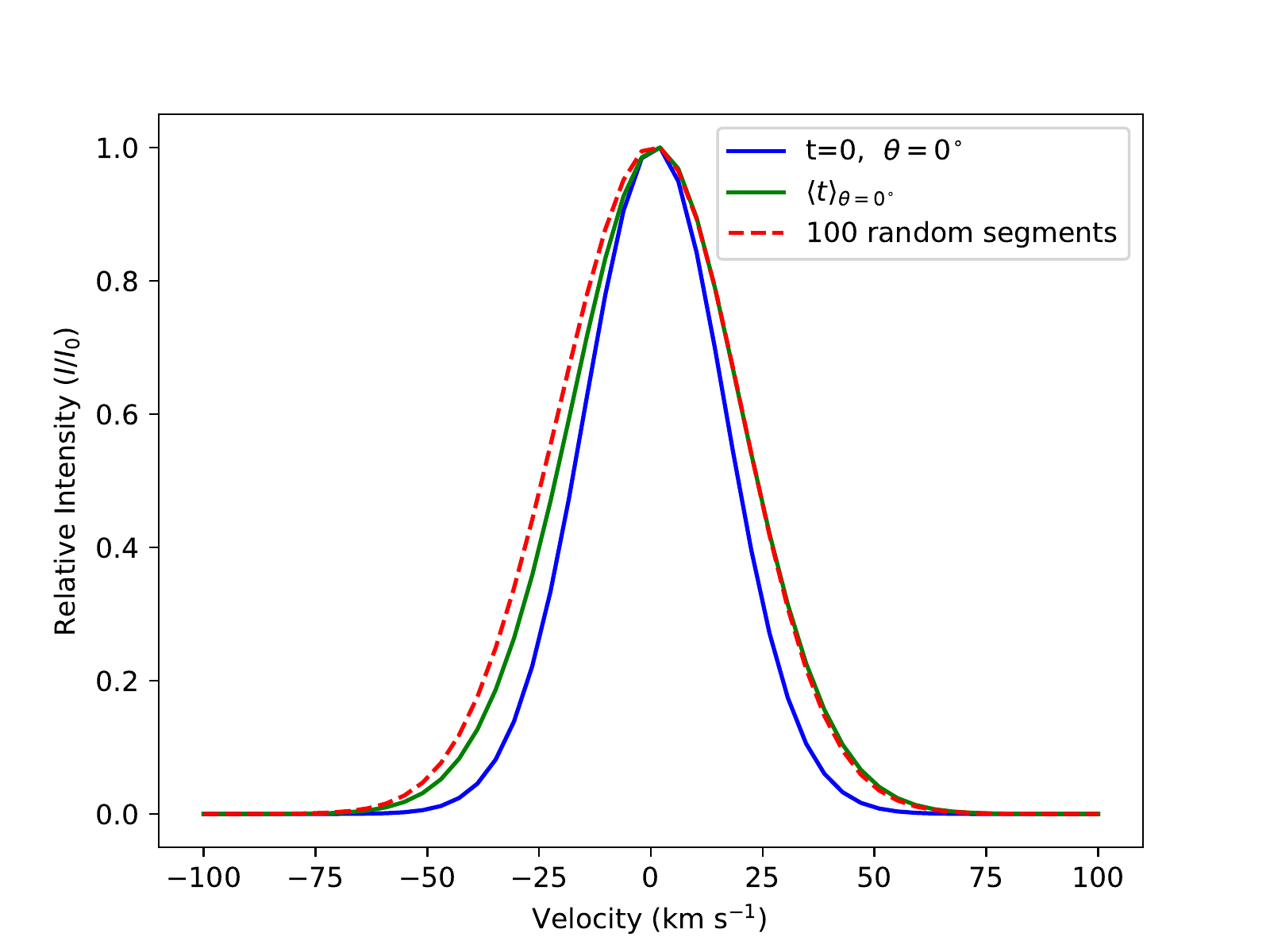}
\includegraphics[scale=0.55]{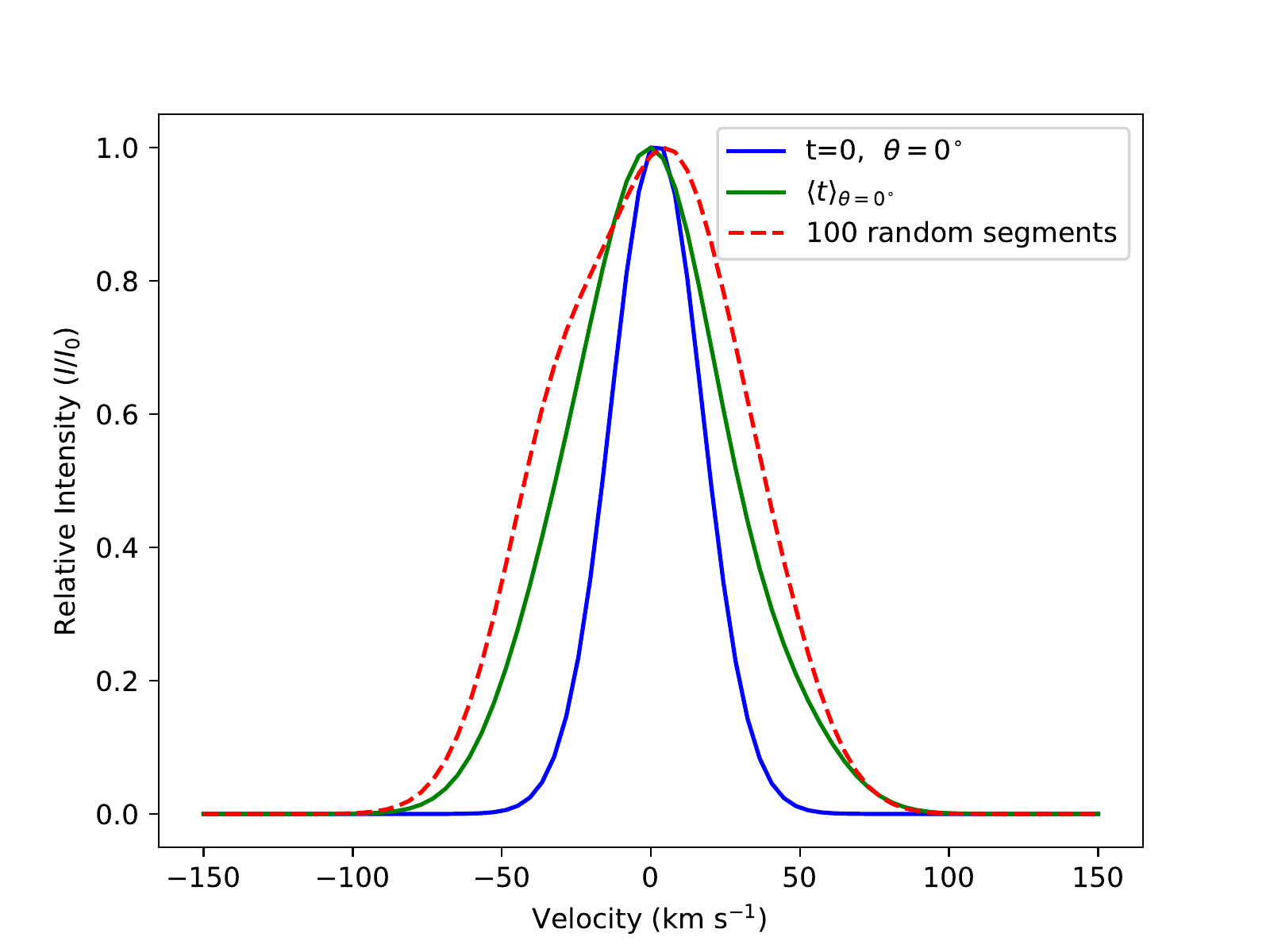}
\caption{{\it Left:} Fe XIII line profile at ($x$,$z$)=(0,-0.9) for LOS=0$^{\circ}$ and $U_{o}$=$\frac{11}{\sqrt{2}}~km~s^{-1}$ at t=0 shown in blue. The line profile averaged over 20 frames is plotted in green. Overplotted in red is the line profile after choosing 100 random segments. {\it Right}: Same as left but for $U_{o}$=$\frac{22}{\sqrt{2}}~km~s^{-1}$. The line profiles in green and red are more broadened compared to the left panel because of the larger wave amplitude.}
\label{fig5}
\end{figure}

\begin{equation}
\begin{aligned}
I_\lambda(x,z',t_{n})=\sum_{j=1}^{100} I_\lambda(x,z'_{j},t_{0,j}+n\delta t);~1\leqslant n \leqslant20,\\
\end{aligned}
\label{eq6}
\end{equation}
where $I_\lambda(,x,z',t_{n})$ is the resultant specific intensity obtained after the superposition of  the intensities of 100 random segments. $LOS_{j}$ (thus $z'_{j}$) is chosen randomly from twelve LOSs; $t_{0,j}$ is the starting time (in frame number) chosen randomly from a uniform distribution of [1,30]. $\delta t$ is the time cadence, kept as 1 frame (or 20 s) in this study. Since $n$ varies from 1-20 frames, the maximum allowed value of the random start time cannot be greater than 30. Following this procedure, the total number of unique segments available in this study are 360 (30$\times$12). One should note, if we increase $n$, the total number of unique segments will decrease because the number of allowed t$_{0,j}$ will decrease. On the other hand, if $n$ is small, very few frames will be available to compute the mean and root mean square (rms) estimates of the non-thermal line widths and Doppler velocities, respectively. 

\subsection{Measured wave properties vs number of segments}
Next, we compute the moments of $I_\lambda(,x,z',t_{n})$ using Equation~\ref{eq5} and derive the Doppler shifts and thus the Doppler velocity, $v_{D}(x,z',t_{n})$, and exponential line width, $\sigma_{1/e}(x,z',t_{n})$, at each instant ($n$) for twenty consecutive frames ($n$=1 -- 20). Then, the non-thermal line widths ($\sigma_{nt}$) are estimated by subtracting quadratically the line width obtained at $t$=0 (22.5 km s$^{-1}$) when no velocity drivers are applied (see Section~\ref{fm}) from the exponential line width. 
Finally, the rms Doppler velocities, $rms~v_{D}(x,z')$, and mean non-thermal line widths, $mean~\sigma_{nt}(x,z')$, are computed over 20 frames using the following relations:
\begin{equation}
\begin{aligned}
rms~v_{D}(x,z')= \sqrt{\frac{\sum_{n=1}^{20}v_{D}(x,z',t_{n})^{2}}{20}},\\
mean~\sigma_{nt}(x,z')=\frac{\sum_{n=1}^{20}\sigma_{nt}(x,z',t_{n})}{20}.
\end{aligned}
\label{eq7}
\end{equation}
It should be noted that the rms Doppler velocities obtained after random sampling of segments is different from the rms velocity of the driver ($v_{rms}$) at the bottom boundary, which is described by Equation~\ref{eq4}.
Additionally, we degrade the spatial solution of the simulation cube to the spatial resolution of the CoMP ($\sim$3200~km) and a similar procedure as outlined above was adopted to estimate rms Doppler velocities and mean non-thermal line widths. 

Given that a segment is obtained by integrating the simulation cube along a LOS perpendicular to the direction of the magnetic field ($x$~axis), a segment is thus integrated at least over $\sim$5~Mm. Therefore, a total of 100 segments are equivalent to a distance of at least 500~Mm, which corresponds to an inclination ($\tan^{-1}(250/R_{sun})$) of 20$^{\circ}$ with respect to the normal to the surface of the Sun. Since the inclination is not large, we do not choose a LOS inclined to the $x$~axis of the simulation cube. This allows us to investigate the variation of the Doppler velocities and line widths with heights without mixing the emission coming from different heights. 

\begin{figure}[ht!]
\centering
\includegraphics[scale=0.55]{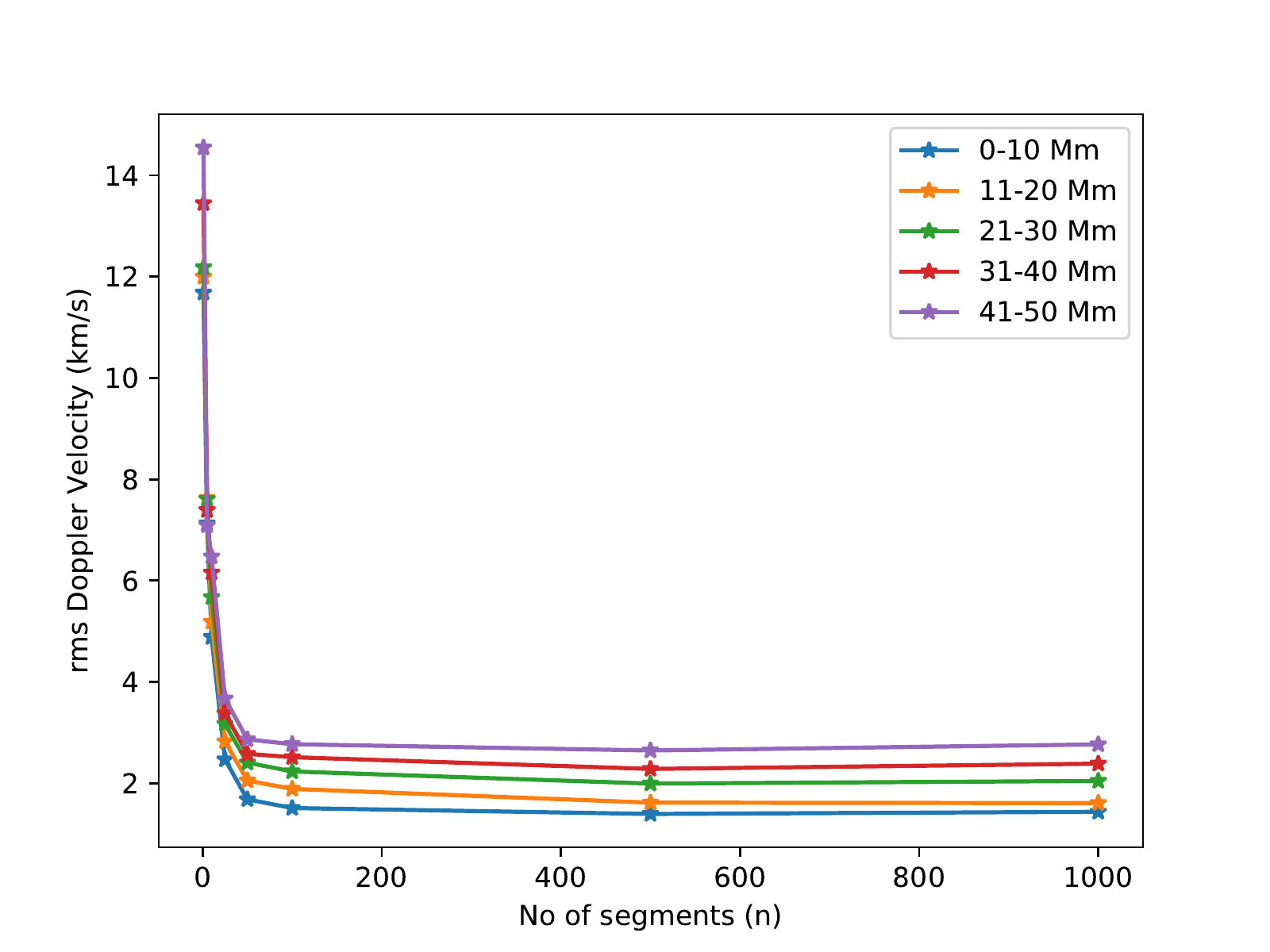}
\includegraphics[scale=0.55]{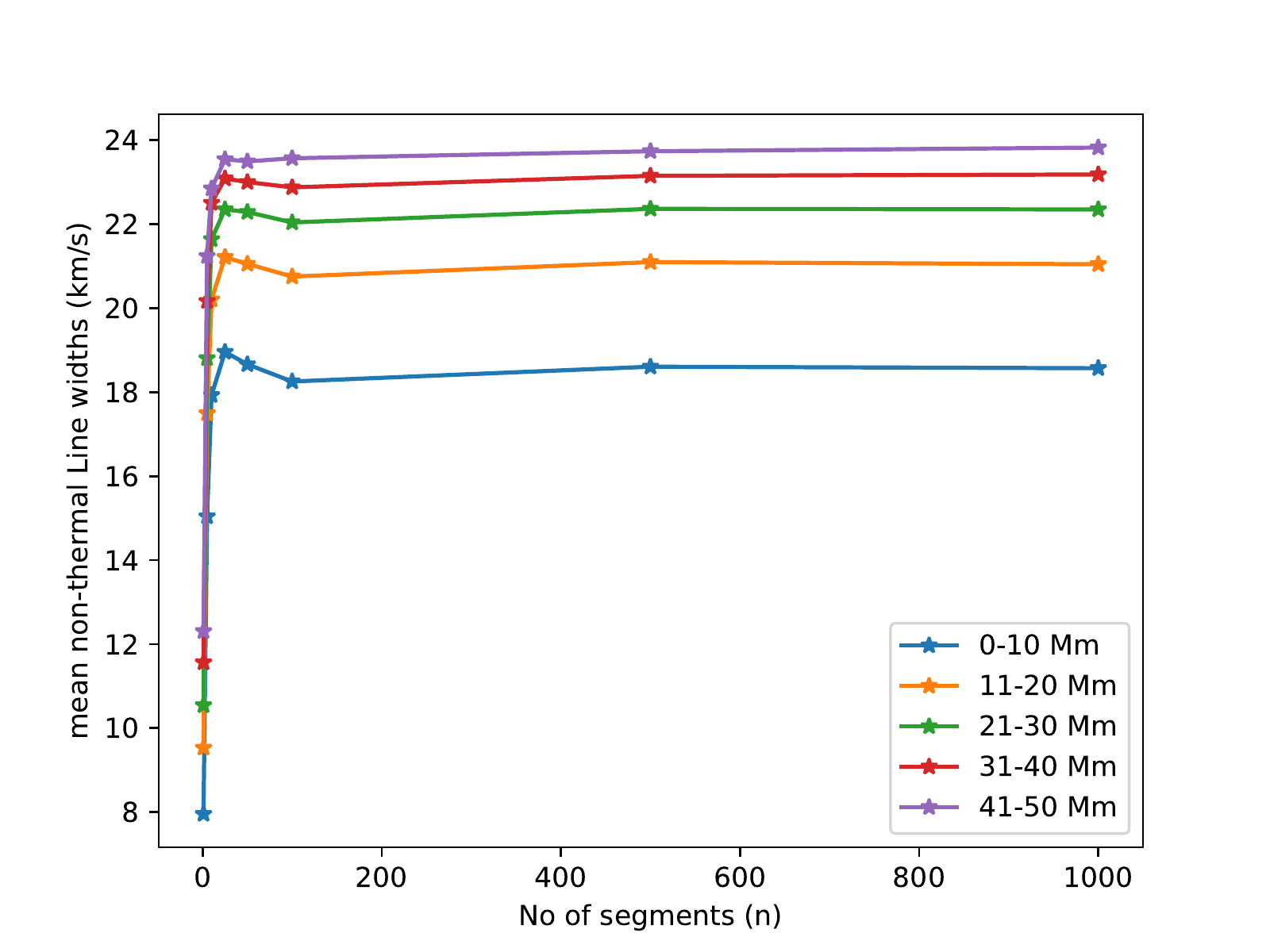}
\caption{{\it Left}: The variation of rms Doppler velocities with total number of random segments for the simulation with $U_{o}$=$\frac{11}{\sqrt{2}}$~km~s$^{-1}$. {\it Right}: Same as left panel but for mean non-thermal line widths.}
\label{fig6}
\end{figure}

\medskip
Furthermore, we also note that the rms Doppler velocities and mean line widths averaged over $y-z$ plane sharply decrease and increase, respectively, with an increase in the number of segments, although both reach a plateau when the number of random segments is greater than 100 (see Figure~\ref{fig6}). Therefore, we choose 100 segments in the current study. This result is partially in agreement with the findings of \citet{2012ApJ...761..138M}, where the authors reported that the rms Doppler velocities monotonically decrease while mean line widths stay constant with increasing number of `threads' (see their Figure~3). This difference may arise due to different methods employed in \citet{2012ApJ...761..138M} and our study. Furthermore, it should be noted that the properties of a `segment' used in this study are different from  a `thread' used in \citet{2012ApJ...761..138M}.  They define a `thread' as an elementary oscillating structure which is uniformly bright and optically thin, while a `segment' used in this study is obtained by integrating the transversely inhomogeneous and gravitationally stratified simulation cube over 5 Mm along a given LOS. Therefore one segment comprises several `thread'-like structures along the LOS but driven uniformly at the bottom boundary. It is worth noting at this point that we do not give any selective weights to the intensities of 100 segments lying along the LOS. In reality, the foreground segments may scatter the photons coming from the background segments. Therefore, the emission from foreground structures might dominate over background structures. This effect is ignored here for the simplicity.\\  


\subsection{Doppler velocity vs non-thermal line width}
\begin{figure}[ht!]
\begin{flushleft}
\includegraphics[scale=0.65,angle=270]{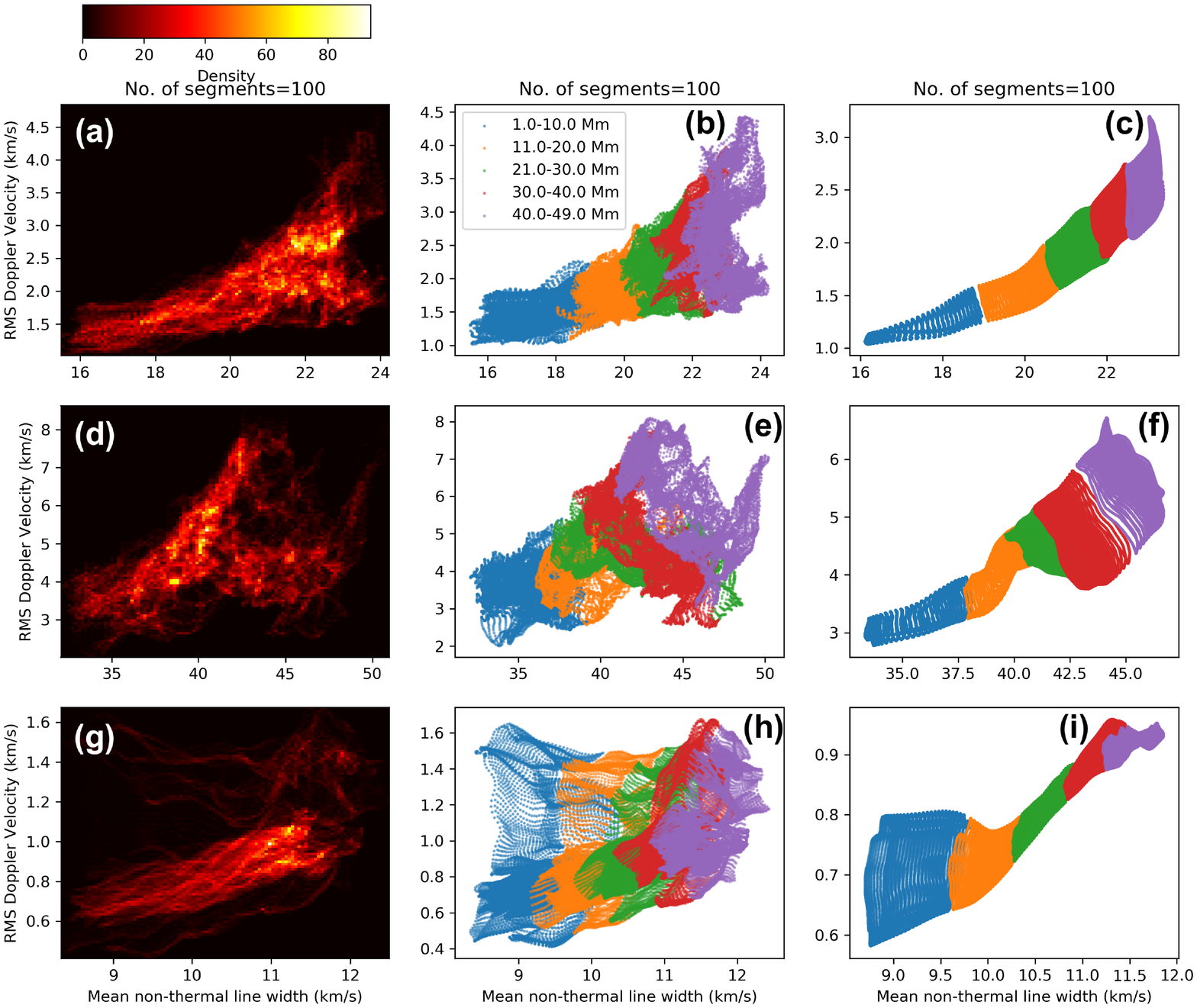}
\caption{(a): Two dimensional density plot representing the variation of the rms Doppler velocity with non-thermal line widths for 100 random segments for the simulation with $U_{o}$=$\frac{11}{\sqrt{2}}$~km~s$^{-1}$. (b): Scatter plot of the variation of the rms Doppler velocities with mean non-thermal line widths for the simulation with $U_{o}$=$\frac{11}{\sqrt{2}}$~km~s$^{-1}$. (c): Same as (b) but degraded for the CoMP spatial resolution ($\sim$3200~km). Different colors represent different height ranges.
(d--f): Same as (a), (b), and (c) but for the simulation with $U_{o}$=$\frac{22}{\sqrt{2}}$~km~s$^{-1}$.  
(g--i): Same as (a), (b), and (c) but for the simulation with $U_{o}$=$\frac{11}{\sqrt{2}}$~km~s$^{-1}$}
\end{flushleft}
\label{fig7}
\end{figure}

To examine relationships between $rms~v_{D}(x,z')$ and $mean~\sigma_{nt}(x,z')$ (Equation~\ref{eq7}), we create density and scatter plots comparing the two variables for all values of $x$ and $z'$. Figure~\ref{fig7} (a) displays a two dimensional density plot of the variation of the rms Doppler velocities and mean non-thermal line widths for the simulation with driver amplitude of $U_{o}$=$\frac{11}{\sqrt{2}}$~km~s$^{-1}$. A wedge shape correlation between rms Doppler velocities and mean non-thermal line widths is evident and is qualitatively similar to Figure~\ref{fig0}. Figure~\ref{fig7} (b) shows the variation of rms Doppler velocities with mean non-thermal line widths segregating different height ranges (shown in different colors) for the same driver amplitude. The mean non-thermal line widths increase with increasing heights, again resembling the observations taken from the CoMP (Figure~\ref{fig0}).  Figure~\ref{fig7} (c) is obtained after degrading the spatial resolution of the simulation cube to the spatial resolution of the CoMP, which is about 65\% of the total extent of the simulation cube.

\medskip

In Figure~\ref{fig7} (c), several Lissajous like curves for a given $x$ can be seen. This is the result of the degradation of the simulation cube. The simulation cube extends up to 5~Mm in $y$ and $z$ directions. Since the periodic boundaries were used in $y$ and $z$ directions while performing simulations, we perform the degradation of the simulation cube by wrapping the $y$ and $z$ boundaries of the simulation cube. This lead to the closed curves (Lissajous curves) in Figure~\ref{fig7} (c) for a particular $x$. We confirmed this effect disappears if, instead of wrapping the cube in $y$ and $z$ directions, nearest-neighbours is used to perform degradation; the curves shown in different colors are no longer closed but open. In addition to this, the degradation of the simulation cube filtered the small spatial variations in the $rms~v_{D}(x,z')$ and $mean~\sigma_{nt}(x,z')$ and only the large spatial variations that are very small compared to the one sigma standard deviations, remained. Such large scale variations in $rms~v_{D}(x,z')$ and $mean~\sigma_{nt}(x,z')$ (due to large scale density variation) for a given $x$ lead to the observed open or closed curves. However, like the real observations, if noise is included, it may distort the observed Lissajous curves and the scatter plots may look more realistic. Such  Lissajous curves do not affect the results of this study and therefore, a wedge-shape correlation can still be noted in Figure~\ref{fig7} (c).

\medskip
The results displayed in Figure~\ref{fig7} (a--c) demonstrate that the mean non-thermal line widths vary from $\sim$15~km~s$^{-1}$ - 24~km~s$^{-1}$.
Though the simulation generates a wedge-shaped correlation (as observed in Figure~\ref{fig0} and \citealp{2012ApJ...761..138M}), the range of the line widths obtained for $v_{rms}$ $\sim$15~km~s$^{-1}$ is narrower than the observed values of non-thermal line widths (which can reach up to 40~km~s$^{-1}$).
Since the non-thermal broadening of emission lines is due to the LOS superposition of different transversely oscillating segments, it is prudent to assume that line widths must depend on the velocity amplitude of these oscillations (see also right panel of Figure~\ref{fig5}). To study the effect of the amplitude of velocity drivers on mean line widths, we perform another simulation with $U_{o}$=$\frac{22}{\sqrt{2}}$~km~s$^{-1}$ that gives a rms velocity of 26~km~s$^{-1}$ at the bottom boundary (obtained using Equation~\ref{eq4}), and the results are shown in Figure~\ref{fig7} (d--f). It can be noted from the figure that the mean non-thermal line widths increase and so does the rms Doppler velocities. This constitutes additional evidence of the correlation between rms Doppler velocities and non-thermal line widths. Moreover, a wedge-shaped correlation similar to Figure~\ref{fig7} (a--c) is also clearly seen. Similarly, we perform another simulation with $U_{o}$=$\frac{5}{\sqrt{2}}$~km~s$^{-1}$ ($v_{rms}$ $\sim$7~km~s$^{-1}$) and found that both the mean non-thermal line widths and rms Doppler velocities decrease (see bottom panels of Figure~\ref{fig7}). 

\begin{figure}[ht!]
\centering
\includegraphics[scale=0.7]{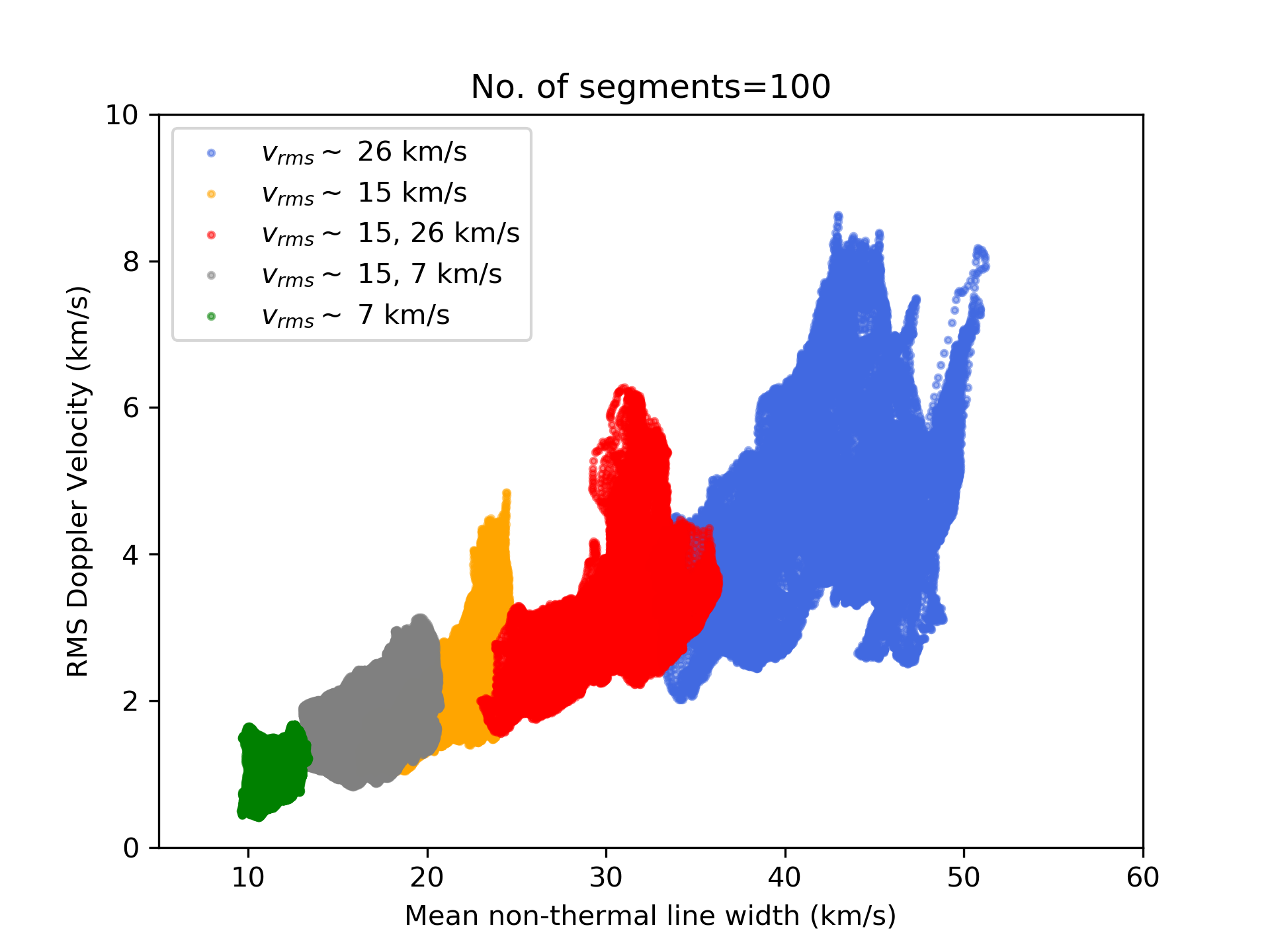}
\caption{Same as figure~\ref{fig7} but including the results obtained by choosing 100 random segments from different velocity drivers shown in different colors.}
\label{fig10}
\end{figure}

Figure~\ref{fig10} presents the variation of the rms Doppler velocities and mean non-thermal line widths for different wave amplitudes of velocity drivers. We also choose random segments from the simulations with different velocity amplitudes as shown in red and grey colors in Figure~\ref{fig10}. The wedge shape correlation can be conspicuously noted. It can be understood from this figure that the non-thermal line widths (and rms Doppler velocities) depend on the input wave amplitudes. Our results are in agreement with \citet{2012ApJ...761..138M}, where the authors have shown that non-thermal broadening of emission line depends on the input wave amplitudes. Further, in our study, the wave amplitudes depend on the density, which decays exponentially with height. Thus transverse wave amplitude increases with height in the corona, and so do the non-thermal line widths. Therefore, the wedge-shape is at least partially due to height-dependent wave amplitude of the transverse MHD waves. Overall, our simulations could reproduce the height-dependent wedge-shape correlation between rms Doppler velocities and mean non-thermal line widths as seen in the observations without artificially adding any additional non-thermal broadening as done in \citet{2012ApJ...761..138M} to match the observed non-thermal line widths. These results allow us to conclude that large non-thermal line widths (due to the unresolved wave amplitudes) in solar corona conceal large wave amplitudes (and hence energies). This relaxes the requirement for artificially adding an extra unknown source of `dark' or `hidden' energy in the solar corona to match the non-thermal line widths.



\subsection{Non-thermal widths vs height}
\begin{figure}[ht!]
\centering
\includegraphics[scale=0.7]{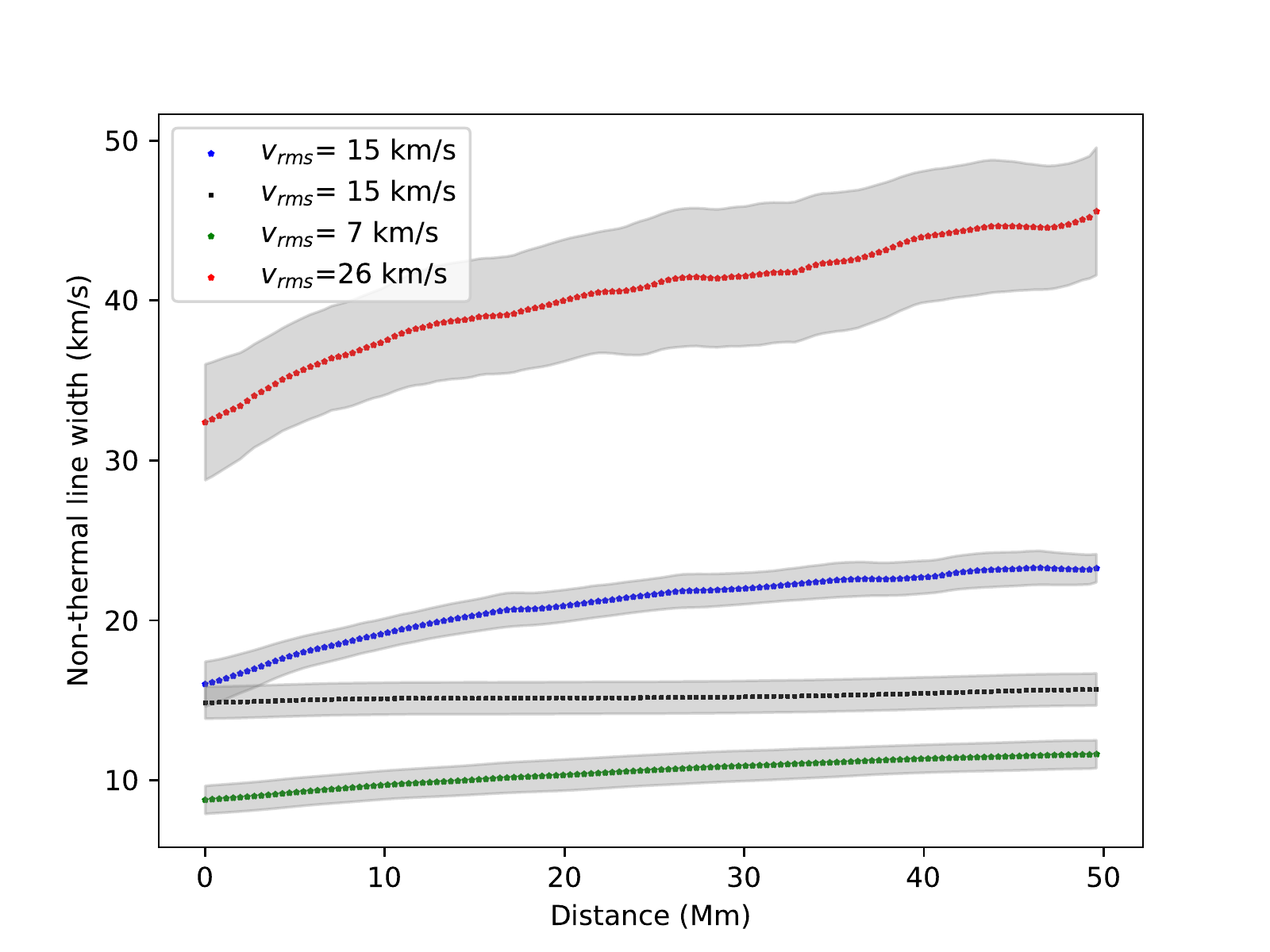}
\caption{Variation of the non-thermal line widths of 100 random segments with height. Different colors represents the results obtained from simulations run with different $v_{rms}$ at the bottom boundary. Blue and black curves shows the variation of the non-thermal line widths for a gravitationally stratified and a uniform plasma along the $x$ axis. Shaded region represents 1-$\sigma$ standard deviation computed over 20 frames.
}
\label{fig11}
\end{figure}

Next, we study the variation of the non-thermal line widths with height above the solar photosphere. Figure~\ref{fig11} demonstrates the increase in the mean non-thermal line widths ($mean~\sigma_{nt}(x,z)$, averaged over $z$ axis for 100 random segments) with height due to the density stratification. We also show this is true for different LOSs and strengths of the velocity drivers ($U_{0}$). Curves in different colors represent the three different velocity drivers. To understand the effects of the gravitational stratification on the line widths, we perform a similar analysis as described above on a simulation (with $U_{o}$=$\frac{11}{\sqrt{2}}$~km~s$^{-1}$) without including gravity. We find that the non-thermal line widths do not vary with height in that case (see the curve in black in Figure~\ref{fig11}). 
Further we note that the non-thermal line widths level-off with increasing heights (Figure~\ref{fig11}). The nature of variation is consistent with the observations of line widths \citep{1990ApJ...348L..77H, 1998A&A...339..208B, 1998SoPh..181...91D, 2009A&A...501L..15B,2012ApJ...753...36H} in coronal holes and a few other MHD simulations \citep[AWSoM;][]{2017ApJ...845...98O}. The levelling-off of the mean line widths could be a signature of wave damping or reflections or both \citep{1998SoPh..181...91D, 2012ApJ...753...36H}. However, the physics behind this plateau is beyond the scope of this paper and will be the subject of a future study. \\

\subsection{Energy estimate}
Finally we estimate the total energy density, $E$, in the simulation cube averaged over the entire duration and volume of the simulation cube. Total energy, $E$ is computed using the following relations.
\begin{equation}
    E(x,y,z,t)=\frac{\rho(x,y,z,t) v^{2}(x,y,z,t)}{2} + \frac{B(x,y,z,t)^2}{2\mu} + \frac{p(x,y,z,t)}{\gamma-1} + \rho(x,y,z,t) \phi(x,y,z,t),
    \label{eq8}
\end{equation}

\begin{equation}
    E=\langle E (t) \rangle_{t}=\langle \frac{1}{V} \int_{V}(E(x,y,z,t) -E(x,y,z,0)) dV' \rangle_{t}.
    \label{eq9}
\end{equation}
Here, $V$ represents the volume of the simulation cube and $\phi$ is the gravitational potential. We compare $E$ derived using equations~\ref{eq8}~and~\ref{eq9} with the time averaged Alfv\'en wave energy density $\langle \rho~rms~v_{D}^2 \rangle_{x,z'}$, estimated using the observed rms Doppler velocities of the synthetic images obtained by random superposition of 100 segments. This expression is frequently used to estimate the energy carried by the transverse and largely incompressible (Alfv\'enic) waves using the observed values of rms Doppler velocity fluctuations \citep{2007Sci...317.1192T,2011Natur.475..477M}. This comparison allow us to estimate the amount of the underestimation of the true wave energies due to the LOS superposition.\\

The average total energy ($E$) of the simulations with $U_{0}$ = $\frac{5}{\sqrt{2}}$~km~s$^{-1}$, $\frac{11}{\sqrt{2}}$~km~s$^{-1}$, and $\frac{22}{\sqrt{2}}$~km~s$^{-1}$ is found to be $\sim$~1.23$\times$10$^{-5}~J~m^{-3}$, $\sim$~4.2$\times$10$^{-5}~J~m^{-3}$, and $\sim$~2.2$\times$10$^{-4}~J~m^{-3}$, respectively. The corresponding `observed' average Alfv\'en wave energy is $\sim$~9$\times$10$^{-8}~J~m^{-3}$, $\sim$~4.1$\times$10$^{-7}~J~m^{-3}$, and $\sim$~2.7$\times$10$^{-6}~J~m^{-3}$. Therefore, the observed Alfv\'en wave energy was found to be $\sim$0.7\%, $\sim$0.9\%, and $\sim$1\%, respectively, of the actual wave energy. This is much less than 10\%-40\% reported in \citet{2012ApJ...746...31D} where the authors considered only ten loops along the LOS integration.\\

We also estimate the time average energy flux, $F$, injected into the simulation domain through the bottom boundary ($x$=0). It can be estimated by the following relation.

\begin{equation}
    F=\langle F \rangle_{t}=\langle \frac{1}{A}\int_{S}E(0,y,z,t)~v_{g}~dA'~\rangle_{t},
    \label{flux}
\end{equation}
where $v_{g}$ is the group speed of the transverse MHD wave which is $\sim$800~km~s$^{-1}$ and $A$ is the area of the bottom boundary.
$F$ at the bottom boundary for the 
simulations with $U_{0}$ = $\frac{5}{\sqrt{2}}$~km~s$^{-1}$, $\frac{11}{\sqrt{2}}$~km~s$^{-1}$, and $\frac{22}{\sqrt{2}}$~km~s$^{-1}$ is found to be $\sim$22~Wm$^{-2}$, $\sim$61~Wm$^{-2}$, and $\sim$262~Wm$^{-2}$ respectively.\\
In addition to this, we also estimate the Poynting flux ($S_{x}=\frac{-1}{\mu_{0}} |({\bf v \times B) \times B} |_{x}$) passing through the bottom boundary of the simulations. Since the simulations are ideal, we have ignored the magnetic diffusivity. However, a small numerical diffusivity will be present in the simulations. $S_{x}$ for the 
simulations with $U_{0}$ = $\frac{5}{\sqrt{2}}$~km~s$^{-1}$, $\frac{11}{\sqrt{2}}$~km~s$^{-1}$, and $\frac{22}{\sqrt{2}}$~km~s$^{-1}$ is found to be $\sim$14~Wm$^{-2}$, $\sim$51~Wm$^{-2}$, and $\sim$213~Wm$^{-2}$ respectively.
Its worth noting that the values of $S_{x}$ are in good agreement with the flux computed using the equation~\ref{flux}.\\
We compare the energy flux computed using equation~\ref{flux} and $S_{x}$ with those estimated using synthetic images obtained by random superposition of 100 segments. It can be calculated by multiplying the time averaged Alfv\'en wave energy density at the bottom boundary ($\langle \rho~rms~v_{D}^2 \rangle_{x,z'}$) with the Alfv\'en speed. The time averaged observed Alfv\'en wave energy flux is estimated to be $\sim$0.04~Wm$^{-2}$, $\sim$0.21~Wm$^{-2}$, and $\sim$1.08~Wm$^{-2}$ for simulations with $U_{0}$ = $\frac{5}{\sqrt{2}}$~km~s$^{-1}$, $\frac{11}{\sqrt{2}}$~km~s$^{-1}$, and $\frac{22}{\sqrt{2}}$~km~s$^{-1}$ respectively. Again, we note that the observed Alfv\'en wave energy is $\sim$0.2\%~-~0.4\%, respectively, of the true wave energy flux. 

Thus our simulations are able to generate large differences of 2-3 orders of magnitude in the synthetically observed and the actual wave energy flux, which had probably tempted \citet{2012ApJ...761..138M} to coin the term `hidden' or `dark' energy.

\section {Discussion and conclusions}
\label{sec5}
A recent debate on the `hidden' or `dark' energy in the solar corona has arisen due to the discrepancy in the wave energies measured in corona using CoMP in comparison to those measured with SDO/AIA. To understand this discrepancy, we have performed a 3D MHD simulation of transverse MHD waves propagating in a gravitationally stratified plasma with a typical setting of coronal holes (open-field regions).

\smallskip
Observational studies using data from the SDO and CoMP have revealed that the polar coronal holes comprise several fine-scale magnetic structures that exhibit periodic transverse motions \citep{2014ApJ...790L...2T,2015NatCo...6E7813M}. Hence, our simulations utilize a series of density inhomogeneities which they are randomly positioned perpendicular to the direction of the magnetic field. Observations of Alfv\'enic waves in these regions revealed a log-normal distribution of periods with a mean value $\sim$ 470~km~s$^{-1}$ and a velocity distribution with a mean value $\sim$~14~km~s$^{-1}$, that can range from $\sim$5~km~s$^{-1}$ - 40~km~s$^{-1}$. In order to model the observed properties of Alfv\'enic waves in coronal holes, a log-normal distribution of period with a mean value of 400 s and velocity drivers with rms wave amplitude of 7~km~s$^{-1}$, 15~km~s$^{-1}$, and 26~km~s$^{-1}$ (with varying phase) are used. 

We find that the forward modelling of Fe XIII 10749~{\AA} emission line leads to a sharp decrease (and increase) in the rms Doppler velocities (and mean line widths) as we integrate over an increasing number of density inhomgenetities. These results are in close agreement with the findings of \citet{2012ApJ...761..138M}. For this study, we perform MHD simulations of propagating transverse MHD wave with three different values of $v_{rms}$ only. More realistic simulations representing the true distribution of wave velocity amplitudes  described in, e.g. \citet{2015NatCo...6E7813M, 2019NatAs...3..223M}, can be performed in the future. Our simulations could generate a wedge-shaped correlation between the rms Doppler velocities and mean non-thermal line widths. We degrade the spatial resolution in our simulations to match the spatial resolution of CoMP but found no significant differences in the rms Doppler velocities and mean non-thermal line widths. In our study, transverse MHD waves and induced uniturbulence are able to produce the observed values of large non-thermal line widths due to LOS integration without including any significant contribution from flows or torsional motions as suggested by \citet{2012ApJ...761..138M,2012ApJ...746...31D}. Further, unlike the method described in \citet{2012ApJ...761..138M}, we have not artificially added any extra non-thermal broadening to match the observed values of non-thermal line widths (15 km s$^{-1}$ - 45 km s$^{-1}$). 
The MHD model used in this study is more detailed because it employs the mechanism of 3D MHD wave propagation, compared to the Monte Carlo model of oscillating `threads' in \citet{2012ApJ...761..138M}.

\citet{2012ApJ...746...31D} have also studied the effects of LOS integrations by employing a 3D MHD simulation. However, the authors have not performed the forward modeling for emission lines and a single pulse of velocity driver was used. In this study, we use a, more realistic, continuous multiple periodic driver with random direction of the polarisation at the bottom boundary to mimic the photospheric driving of the plumes. 

There are a number of caveats that come with our results.
As seen by comparing Figure~\ref{fig0} (and previous studies, \citealp{2007Sci...317.1192T,2012ApJ...761..138M, 2015NatCo...6E7813M}) to Figure~\ref{fig10}, the rms Doppler velocities calculated from our synthetic emission lines are larger than those observed in the CoMP data. One possible reason could be that the CoMP instrument has a finite exposure time. In our simulations, a snapshot is recorded instantaneously (exposure time $\sim$ 0 s). The finite exposure time leads to the averaging of spectra in time, thus reducing the magnitude of measured Doppler velocity fluctuations (as occurs from spatial averaging). Unfortunately we are unable to test this with the current simulations as the available unique segments are limited by the total time of the simulation (which is in turn limited by the computational resources and time). 

Moreover, although we can reproduce the wedge-shape of the joint distribution of rms Doppler Velocity and non-thermal lines widths, it should be borne in mind that here we give a proof of concept that focuses on transverse waves propagating in open field regions. The exact shape of the correlation will depend on the conditions of the solar plasma at the time of observations which is of significantly greater complexity to model and well beyond the scope of this study.\\

The propagating waves in this study suffers a weak damping by resonant absorption because the simulation domain (50 Mm) is smaller than the damping length ($\sim$ 200 Mm) which is comparable to the wavelength of the transverse (kink) wave \citep{2010ApJ...711..990P}. This is further confirmed by performing a simulation without any gravitational stratification, where line widths did not decrease with height.  \citet{2017NatSR...714820M} have also reported that the rms wave amplitude and  Alfv\'enic wave energy flux in their MHD simulation without the inclusion of gravity were weakly damped over 50 Mm.
Furthermore, we find that the non-thermal line widths first increase and then level off with the height in our simulations. \citet{2012ApJ...753...36H} investigated several different mechanisms such as the effect of scattered light, photoexcitation of the emission lines, and inhomogeneities in the temperatures to explain the levelling-off and decrease in the non-thermal line widths. These authors finally proposed that the damping of waves can explain the observed decrease in the line widths. In our study, we rule out significant wave damping due to resonant absorption, due to numerical viscosity, and resistivity because if these were the dominant wave damping mechanisms then a decrease in the non-thermal line widths must have been seen in the simulation without the gravitational stratification. For a similar reason, we rule out that a substantial energy of the waves was spent in ohmic heating of the plasma \citep[also][]{2017NatSR...714820M}. Moreover the non-thermal line widths in the non-stratified MHD simulation increase by 1-2 km s$^{-1}$ with increasing height. It allows us to conclude that the induced uniturbulence in these simulations does not play a significant role in the non-thermal broadening. Unresolved wave amplitudes are the main reason for the non-thermal broadening of the optically thin emission lines. Apart from the damping, reflections of wave due to density stratification is also expected in our simulation. The reflection of Alfv\'en wave energy due to gravitational stratification leading to the levelling-off non-thermal line widths is reported in \citet{2017ApJ...845...98O}. The detailed mechanism of wave reflection and its effect on the variation of the non-thermal line widths in our simulations is beyond the focus of the current study and will be explored in the future.


Our study indicates that a spectrograph with good spectral resolution will not be able to resolve the Doppler velocities due to LOS superposition of the emission spectra of different structures, which is the result of the optically thin nature of the solar corona. However, a good spatial resolution may be useful to resolve the POS motions (similar to SDO/AIA). CoMP suffers from both coarse spatial resolution and the LOS superposition of structures in the optically thin corona. Therefore CoMP could not resolve the true wave amplitudes in the LOS Doppler velocity fluctuations and POS velocity fluctuations. This lead to the gross underestimation of the true wave energy flux. Our study reveals that only 0.2-1\% of true wave energy flux or energy density is estimated by the resolved Doppler velocity fluctuations. The unresolved wave amplitudes results in the non-thermal broadening. Therefore, we conclude that Doppler velocities are not the true representatives of the wave amplitudes and wave energies. True wave energies are hidden (in form of unresolved wave amplitudes) in the non-thermal line widths. Our study has revealed the sites of `hidden' or `dark' wave energies without adding any unknown source of artificial energy. Finally, to estimate the energy budget and to explain the heating of the solar corona, a relation between true wave amplitudes and the non-thermal line width is needed and will be explored in the future.

\appendix
\section{Photoexcitation of Fe XIII 10749~{\AA}}
 The electron transitions for the Fe XIII line become dominated by photoionisation at a certain height above the solar photosphere. The specific intensity of a spectral line depends on the rate of the collisional excitations and radiative excitations (or photoionisation). The collisional excitation rate varies as $n_{e}^{2}$, while radiative excitation rate varies as $n_{e}$ \citep{2003ApJS..144..135Y,2016JGRA..121.8237L}. Therefore, in the low corona, emission of the Fe XIII is dominated by the collisional excitations. While at the greater heights, the transitions are dominated by radiative excitations. Usually the contribution of the radiative excitations are low for typically use coronal emission lines, which are in the EUV, but for the infrared Fe XIII their contribution becomes significant at distances as low as 1.1 $R_{\odot}$ \citep{2016JGRA..121.8237L,2018ApJ...852...52D}.

 In order to examine whether this influences our results, the forward modeling was also performed including the contribution of photoionisation where the source is set at a height of $1R_{\odot}$. The photoionisation is incorporated by including the rates of radiative excitation in addition to collisional excitations while computing the ionic fraction of the Fe XIII emission line and $G (n_{e},T)$. We assume that photoionisation happens due to a background radiation field emitted by a blackbody at $T$=5700~K. The detailed mechanism of computing $G$ including photoionisation can be found in \citet{2003ApJS..144..135Y}. The results presented in the proceeding sections are identical whether photoionisation is included or not. This is likely due to our simulations extended up to 1.07~$R_{\odot}$, which is less than one pressure scale height. Hence, the electron density does not become so small that photoionisation dominates over collisional excitations. Thus, throughout the manuscript, we have not included the effects of photoionisation in our calculations of synthetic radiation from the Fe XIII emission line.
 
\acknowledgements
We thank the anonymous referee for useful suggestions that have improved the manuscript. TVD and VP were supported by the GOA-2015-014 (KU~Leuven) and the European Research Council (ERC) under the European Union's Horizon 2020 research and innovation programme (grant agreement No 724326). We also acknowledge Dr. Hui Tian and Prof. Valery Nakariakov for insightful discussions. The authors acknowledge the work of the National Center for Atmospheric Research/High Altitude Observatory CoMP instrument team.



\begin{thebibliography}{}
\expandafter\ifx\csname natexlab\endcsname\relax\def\natexlab#1{#1}\fi
\providecommand{\url}[1]{\href{#1}{#1}}
\providecommand{\dodoi}[1]{doi:~\href{http://doi.org/#1}{\nolinkurl{#1}}}
\providecommand{\doeprint}[1]{\href{http://ascl.net/#1}{\nolinkurl{http://ascl.net/#1}}}
\providecommand{\doarXiv}[1]{\href{https://arxiv.org/abs/#1}{\nolinkurl{https://arxiv.org/abs/#1}}}

\bibitem[{{Abbo} {et~al.}(2016){Abbo}, {Ofman}, {Antiochos}, {Hansteen},
  {Harra}, {Ko}, {Lapenta}, {Li}, {Riley}, {Strachan}, {von Steiger}, \&
  {Wang}}]{2016SSRv..201...55A}
{Abbo}, L., {Ofman}, L., {Antiochos}, S.~K., {et~al.} 2016, \ssr, 201, 55,
  \dodoi{10.1007/s11214-016-0264-1}

\bibitem[{{Anfinogentov} {et~al.}(2013){Anfinogentov}, {Nistic{\`o}}, \&
  {Nakariakov}}]{2013A&A...560A.107A}
{Anfinogentov}, S., {Nistic{\`o}}, G., \& {Nakariakov}, V.~M. 2013, \aap, 560,
  A107, \dodoi{10.1051/0004-6361/201322094}

\bibitem[{{Antolin} \& {Van Doorsselaere}(2013)}]{refId0}
{Antolin}, \& {Van Doorsselaere}. 2013, A\&A, 555, A74,
  \dodoi{10.1051/0004-6361/201220784}

\bibitem[{{Arregui}(2015)}]{2015RSPTA.37340261A}
{Arregui}, I. 2015, Philosophical Transactions of the Royal Society of London
  Series A, 373, 20140261, \dodoi{10.1098/rsta.2014.0261}

\bibitem[{{Aschwanden} {et~al.}(2002){Aschwanden}, {de Pontieu}, {Schrijver},
  \& {Title}}]{2002SoPh..206...99A}
{Aschwanden}, M.~J., {de Pontieu}, B., {Schrijver}, C.~J., \& {Title}, A.~M.
  2002, \solphys, 206, 99, \dodoi{10.1023/A:1014916701283}

\bibitem[{{Banerjee} {et~al.}(2007){Banerjee}, {Erd{\'e}lyi}, {Oliver}, \&
  {O'Shea}}]{2007SoPh..246....3B}
{Banerjee}, D., {Erd{\'e}lyi}, R., {Oliver}, R., \& {O'Shea}, E. 2007,
  \solphys, 246, 3, \dodoi{10.1007/s11207-007-9029-z}

\bibitem[{{Banerjee} {et~al.}(2009){Banerjee}, {P{\'e}rez-Su{\'a}rez}, \&
  {Doyle}}]{2009A&A...501L..15B}
{Banerjee}, D., {P{\'e}rez-Su{\'a}rez}, D., \& {Doyle}, J.~G. 2009, \aap, 501,
  L15, \dodoi{10.1051/0004-6361/200912242}

\bibitem[{{Banerjee} {et~al.}(1998){Banerjee}, {Teriaca}, {Doyle}, \&
  {Wilhelm}}]{1998A&A...339..208B}
{Banerjee}, D., {Teriaca}, L., {Doyle}, J.~G., \& {Wilhelm}, K. 1998, \aap,
  339, 208

\bibitem[{{Cranmer} {et~al.}(2007){Cranmer}, {van Ballegooijen}, \&
  {Edgar}}]{2007ApJS..171..520C}
{Cranmer}, S.~R., {van Ballegooijen}, A.~A., \& {Edgar}, R.~J. 2007, \apjs,
  171, 520, \dodoi{10.1086/518001}

\bibitem[{{Cranmer} \& {Winebarger}(2018)}]{2018arXiv181100461C}
{Cranmer}, S.~R., \& {Winebarger}, A.~R. 2018, arXiv e-prints.
\newblock \doarXiv{1811.00461}

\bibitem[{{De Moortel} \& {Pascoe}(2012)}]{2012ApJ...746...31D}
{De Moortel}, I., \& {Pascoe}, D.~J. 2012, \apj, 746, 31,
  \dodoi{10.1088/0004-637X/746/1/31}

\bibitem[{{De Pontieu} {et~al.}(2007){De Pontieu}, {McIntosh}, {Carlsson},
  {Hansteen}, {Tarbell}, {Schrijver}, {Title}, {Shine}, {Tsuneta}, {Katsukawa},
  {Ichimoto}, {Suematsu}, {Shimizu}, \& {Nagata}}]{2007Sci...318.1574D}
{De Pontieu}, B., {McIntosh}, S.~W., {Carlsson}, M., {et~al.} 2007, Science,
  318, 1574, \dodoi{10.1126/science.1151747}

\bibitem[{{Del Zanna} \& {DeLuca}(2018)}]{2018ApJ...852...52D}
{Del Zanna}, G., \& {DeLuca}, E.~E. 2018, \apj, 852, 52,
  \dodoi{10.3847/1538-4357/aa9edf}

\bibitem[{{Dere} {et~al.}(1997){Dere}, {Landi}, {Mason}, {Monsignori Fossi}, \&
  {Young}}]{1997A&AS..125..149D}
{Dere}, K.~P., {Landi}, E., {Mason}, H.~E., {Monsignori Fossi}, B.~C., \&
  {Young}, P.~R. 1997, \aaps, 125, 149, \dodoi{10.1051/aas:1997368}

\bibitem[{{Doschek} {et~al.}(1976{\natexlab{a}}){Doschek}, {Feldman}, \&
  {Bohlin}}]{1976ApJ...205L.177D}
{Doschek}, G.~A., {Feldman}, U., \& {Bohlin}, J.~D. 1976{\natexlab{a}}, \apjl,
  205, L177, \dodoi{10.1086/182118}

\bibitem[{{Doschek} {et~al.}(1976{\natexlab{b}}){Doschek}, {Feldman},
  {Vanhoosier}, \& {Bartoe}}]{1976ApJS...31..417D}
{Doschek}, G.~A., {Feldman}, U., {Vanhoosier}, M.~E., \& {Bartoe}, J.-D.~F.
  1976{\natexlab{b}}, \apjs, 31, 417, \dodoi{10.1086/190386}

\bibitem[{{Doschek} {et~al.}(2007){Doschek}, {Mariska}, {Warren}, {Brown},
  {Culhane}, {Hara}, {Watanabe}, {Young}, \& {Mason}}]{2007ApJ...667L.109D}
{Doschek}, G.~A., {Mariska}, J.~T., {Warren}, H.~P., {et~al.} 2007, \apjl, 667,
  L109, \dodoi{10.1086/522087}

\bibitem[{{Doyle} {et~al.}(1998){Doyle}, {Banerjee}, \&
  {Perez}}]{1998SoPh..181...91D}
{Doyle}, J.~G., {Banerjee}, D., \& {Perez}, M.~E. 1998, \solphys, 181, 91,
  \dodoi{10.1023/A:1005019931323}

\bibitem[{{Feldman} {et~al.}(1976){Feldman}, {Doschek}, {Vanhoosier}, \&
  {Purcell}}]{1976ApJS...31..445F}
{Feldman}, U., {Doschek}, G.~A., {Vanhoosier}, M.~E., \& {Purcell}, J.~D. 1976,
  \apjs, 31, 445, \dodoi{10.1086/190387}

\bibitem[{{Ferraro} \& {Plumpton}(1958)}]{1958ApJ...127..459F}
{Ferraro}, C.~A., \& {Plumpton}, C. 1958, \apj, 127, 459,
  \dodoi{10.1086/146474}

\bibitem[{{Goossens} {et~al.}(2012){Goossens}, {Andries}, {Soler}, {Van
  Doorsselaere}, {Arregui}, \& {Terradas}}]{2012ApJ...753..111G}
{Goossens}, M., {Andries}, J., {Soler}, R., {et~al.} 2012, \apj, 753, 111,
  \dodoi{10.1088/0004-637X/753/2/111}

\bibitem[{{Goossens} {et~al.}(2009){Goossens}, {Terradas}, {Andries},
  {Arregui}, \& {Ballester}}]{2009A&A...503..213G}
{Goossens}, M., {Terradas}, J., {Andries}, J., {Arregui}, I., \& {Ballester},
  J.~L. 2009, \aap, 503, 213, \dodoi{10.1051/0004-6361/200912399}

\bibitem[{{Hahn} {et~al.}(2012){Hahn}, {Landi}, \&
  {Savin}}]{2012ApJ...753...36H}
{Hahn}, M., {Landi}, E., \& {Savin}, D.~W. 2012, \apj, 753, 36,
  \dodoi{10.1088/0004-637X/753/1/36}

\bibitem[{{Hassler} {et~al.}(1990){Hassler}, {Rottman}, {Shoub}, \&
  {Holzer}}]{1990ApJ...348L..77H}
{Hassler}, D.~M., {Rottman}, G.~J., {Shoub}, E.~C., \& {Holzer}, T.~E. 1990,
  \apjl, 348, L77, \dodoi{10.1086/185635}

\bibitem[{{Hollweg}(1973)}]{1973ApJ...181..547H}
{Hollweg}, J.~V. 1973, \apj, 181, 547, \dodoi{10.1086/152072}

\bibitem[{{Hollweg}(1978)}]{1978SoPh...56..305H}
---. 1978, \solphys, 56, 305, \dodoi{10.1007/BF00152474}

\bibitem[{{Hollweg}(1990)}]{1990CoPhR..12..205H}
---. 1990, Computer Physics Reports, 12, 205,
  \dodoi{10.1016/0167-7977(90)90011-T}

\bibitem[{{Karampelas} {et~al.}(2017){Karampelas}, {Van Doorsselaere}, \&
  {Antolin}}]{2017A&A...604A.130K}
{Karampelas}, K., {Van Doorsselaere}, T., \& {Antolin}, P. 2017, \aap, 604,
  A130, \dodoi{10.1051/0004-6361/201730598}

\bibitem[{{Karampelas} {et~al.}(2019){Karampelas}, {Van Doorsselaere}, \&
  {Guo}}]{2019arXiv190102676K}
{Karampelas}, K., {Van Doorsselaere}, T., \& {Guo}, M. 2019, arXiv e-prints.
\newblock \doarXiv{1901.02676}

\bibitem[{{Klimchuk}(2006)}]{2006SoPh..234...41K}
{Klimchuk}, J.~A. 2006, \solphys, 234, 41, \dodoi{10.1007/s11207-006-0055-z}

\bibitem[{{Kohl} {et~al.}(1999){Kohl}, {Esser}, {Cranmer}, {Fineschi},
  {Gardner}, {Panasyuk}, {Strachan}, {Suleiman}, {Frazin}, \&
  {Noci}}]{1999ApJ...510L..59K}
{Kohl}, J.~L., {Esser}, R., {Cranmer}, S.~R., {et~al.} 1999, \apjl, 510, L59,
  \dodoi{10.1086/311793}

\bibitem[{{Landi} {et~al.}(2016){Landi}, {Habbal}, \&
  {Tomczyk}}]{2016JGRA..121.8237L}
{Landi}, E., {Habbal}, S.~R., \& {Tomczyk}, S. 2016, Journal of Geophysical
  Research (Space Physics), 121, 8237, \dodoi{10.1002/2016JA022598}

\bibitem[{{Lau} \& {Siregar}(1996)}]{1996ApJ...465..451L}
{Lau}, Y.-T., \& {Siregar}, E. 1996, \apj, 465, 451, \dodoi{10.1086/177432}

\bibitem[{{Magyar} \& {Van Doorsselaere}(2018)}]{2018ApJ...856..144M}
{Magyar}, N., \& {Van Doorsselaere}, T. 2018, \apj, 856, 144,
  \dodoi{10.3847/1538-4357/aab42c}

\bibitem[{{Magyar} {et~al.}(2017){Magyar}, {Van Doorsselaere}, \&
  {Goossens}}]{2017NatSR...714820M}
{Magyar}, N., {Van Doorsselaere}, T., \& {Goossens}, M. 2017, Scientific
  Reports, 7, 14820, \dodoi{10.1038/s41598-017-13660-1}

\bibitem[{{Matthaeus} {et~al.}(1999){Matthaeus}, {Zank}, {Oughton}, {Mullan},
  \& {Dmitruk}}]{1999ApJ...523L..93M}
{Matthaeus}, W.~H., {Zank}, G.~P., {Oughton}, S., {Mullan}, D.~J., \&
  {Dmitruk}, P. 1999, \apjl, 523, L93, \dodoi{10.1086/312259}

\bibitem[{{McIntosh} \& {De Pontieu}(2012)}]{2012ApJ...761..138M}
{McIntosh}, S.~W., \& {De Pontieu}, B. 2012, \apj, 761, 138,
  \dodoi{10.1088/0004-637X/761/2/138}

\bibitem[{{McIntosh} {et~al.}(2011){McIntosh}, {de Pontieu}, {Carlsson},
  {Hansteen}, {Boerner}, \& {Goossens}}]{2011Natur.475..477M}
{McIntosh}, S.~W., {de Pontieu}, B., {Carlsson}, M., {et~al.} 2011, \nat, 475,
  477, \dodoi{10.1038/nature10235}

\bibitem[{{Morton} {et~al.}(2015){Morton}, {Tomczyk}, \&
  {Pinto}}]{2015NatCo...6E7813M}
{Morton}, R.~J., {Tomczyk}, S., \& {Pinto}, R. 2015, Nature Communications, 6,
  7813, \dodoi{10.1038/ncomms8813}

\bibitem[{{Morton} {et~al.}(2016){Morton}, {Tomczyk}, \&
  {Pinto}}]{2016ApJ...828...89M}
{Morton}, R.~J., {Tomczyk}, S., \& {Pinto}, R.~F. 2016, \apj, 828, 89,
  \dodoi{10.3847/0004-637X/828/2/89}

\bibitem[{{Morton} {et~al.}(2019){Morton}, {Weberg}, \&
  {McLaughlin}}]{2019NatAs...3..223M}
{Morton}, R.~J., {Weberg}, M.~J., \& {McLaughlin}, J.~A. 2019, Nature
  Astronomy, 196, \dodoi{10.1038/s41550-018-0668-9}

\bibitem[{{Nakariakov} {et~al.}(1999){Nakariakov}, {Ofman}, {Deluca},
  {Roberts}, \& {Davila}}]{1999Sci...285..862N}
{Nakariakov}, V.~M., {Ofman}, L., {Deluca}, E.~E., {Roberts}, B., \& {Davila},
  J.~M. 1999, Science, 285, 862, \dodoi{10.1126/science.285.5429.862}

\bibitem[{{Nistic{\`o}} {et~al.}(2013){Nistic{\`o}}, {Nakariakov}, \&
  {Verwichte}}]{2013A&A...552A..57N}
{Nistic{\`o}}, G., {Nakariakov}, V.~M., \& {Verwichte}, E. 2013, \aap, 552,
  A57, \dodoi{10.1051/0004-6361/201220676}

\bibitem[{{Ofman} \& {Davila}(1997{\natexlab{a}})}]{1997ApJ...476L..51O}
{Ofman}, L., \& {Davila}, J.~M. 1997{\natexlab{a}}, \apjl, 476, L51,
  \dodoi{10.1086/310491}

\bibitem[{{Ofman} \& {Davila}(1997{\natexlab{b}})}]{1997ApJ...476..357O}
---. 1997{\natexlab{b}}, \apj, 476, 357, \dodoi{10.1086/303603}

\bibitem[{{Ofman} \& {Davila}(1998)}]{1998JGR...10323677O}
---. 1998, \jgr, 103, 23677, \dodoi{10.1029/98JA01996}

\bibitem[{{Oran} {et~al.}(2017){Oran}, {Landi}, {van der Holst}, {Sokolov}, \&
  {Gombosi}}]{2017ApJ...845...98O}
{Oran}, R., {Landi}, E., {van der Holst}, B., {Sokolov}, I.~V., \& {Gombosi},
  T.~I. 2017, \apj, 845, 98, \dodoi{10.3847/1538-4357/aa7fec}

\bibitem[{{Oran} {et~al.}(2013){Oran}, {van der Holst}, {Landi}, {Jin},
  {Sokolov}, \& {Gombosi}}]{2013ApJ...778..176O}
{Oran}, R., {van der Holst}, B., {Landi}, E., {et~al.} 2013, \apj, 778, 176,
  \dodoi{10.1088/0004-637X/778/2/176}

\bibitem[{{Orta} {et~al.}(2003){Orta}, {Huerta}, \&
  {Boynton}}]{2003ApJ...596..646O}
{Orta}, J.~A., {Huerta}, M.~A., \& {Boynton}, G.~C. 2003, \apj, 596, 646,
  \dodoi{10.1086/377706}

\bibitem[{{O'Shea} {et~al.}(2005){O'Shea}, {Banerjee}, \&
  {Doyle}}]{2005A&A...436L..35O}
{O'Shea}, E., {Banerjee}, D., \& {Doyle}, J.~G. 2005, \aap, 436, L35,
  \dodoi{10.1051/0004-6361:200500120}

\bibitem[{{Parnell} \& {De Moortel}(2012)}]{2012RSPTA.370.3217P}
{Parnell}, C.~E., \& {De Moortel}, I. 2012, Philosophical Transactions of the
  Royal Society of London Series A, 370, 3217, \dodoi{10.1098/rsta.2012.0113}

\bibitem[{{Pascoe} {et~al.}(2010){Pascoe}, {Wright}, \& {De
  Moortel}}]{2010ApJ...711..990P}
{Pascoe}, D.~J., {Wright}, A.~N., \& {De Moortel}, I. 2010, \apj, 711, 990,
  \dodoi{10.1088/0004-637X/711/2/990}

\bibitem[{{Porth} {et~al.}(2014){Porth}, {Xia}, {Hendrix}, {Moschou}, \&
  {Keppens}}]{2014ApJS..214....4P}
{Porth}, O., {Xia}, C., {Hendrix}, T., {Moschou}, S.~P., \& {Keppens}, R. 2014,
  \apjs, 214, 4, \dodoi{10.1088/0067-0049/214/1/4}

\bibitem[{{Suzuki} \& {Inutsuka}(2006)}]{2006JGRA..111.6101S}
{Suzuki}, T.~K., \& {Inutsuka}, S.-I. 2006, Journal of Geophysical Research
  (Space Physics), 111, A06101, \dodoi{10.1029/2005JA011502}

\bibitem[{{Thurgood} {et~al.}(2014){Thurgood}, {Morton}, \&
  {McLaughlin}}]{2014ApJ...790L...2T}
{Thurgood}, J.~O., {Morton}, R.~J., \& {McLaughlin}, J.~A. 2014, \apjl, 790,
  L2, \dodoi{10.1088/2041-8205/790/1/L2}

\bibitem[{{Tiwari} {et~al.}(2019){Tiwari}, {Morton}, \&
  {McLaughlin}}]{Tiwari2019}
{Tiwari}, A., {Morton}, R.~J., \& {McLaughlin}, J.~A. 2019, ApJ

\bibitem[{{Tomczyk} \& {McIntosh}(2009)}]{2009ApJ...697.1384T}
{Tomczyk}, S., \& {McIntosh}, S.~W. 2009, \apj, 697, 1384,
  \dodoi{10.1088/0004-637X/697/2/1384}

\bibitem[{{Tomczyk} {et~al.}(2007){Tomczyk}, {McIntosh}, {Keil}, {Judge},
  {Schad}, {Seeley}, \& {Edmondson}}]{2007Sci...317.1192T}
{Tomczyk}, S., {McIntosh}, S.~W., {Keil}, S.~L., {et~al.} 2007, Science, 317,
  1192, \dodoi{10.1126/science.1143304}

\bibitem[{{Tomczyk} {et~al.}(2008){Tomczyk}, {Card}, {Darnell}, {Elmore},
  {Lull}, {Nelson}, {Streander}, {Burkepile}, {Casini}, \&
  {Judge}}]{2008SoPh..247..411T}
{Tomczyk}, S., {Card}, G.~L., {Darnell}, T., {et~al.} 2008, \solphys, 247, 411,
  \dodoi{10.1007/s11207-007-9103-6}

\bibitem[{{van Ballegooijen} {et~al.}(2011){van Ballegooijen}, {Asgari-Targhi},
  {Cranmer}, \& {DeLuca}}]{2011ApJ...736....3V}
{van Ballegooijen}, A.~A., {Asgari-Targhi}, M., {Cranmer}, S.~R., \& {DeLuca},
  E.~E. 2011, \apj, 736, 3, \dodoi{10.1088/0004-637X/736/1/3}

\bibitem[{{van Ballegooijen} {et~al.}(2017){van Ballegooijen}, {Asgari-Targhi},
  \& {Voss}}]{2017ApJ...849...46V}
{van Ballegooijen}, A.~A., {Asgari-Targhi}, M., \& {Voss}, A. 2017, \apj, 849,
  46, \dodoi{10.3847/1538-4357/aa9118}

\bibitem[{{van der Holst} {et~al.}(2014){van der Holst}, {Sokolov}, {Meng},
  {Jin}, {Manchester}, {T{\'o}th}, \& {Gombosi}}]{2014ApJ...782...81V}
{van der Holst}, B., {Sokolov}, I.~V., {Meng}, X., {et~al.} 2014, \apj, 782,
  81, \dodoi{10.1088/0004-637X/782/2/81}

\bibitem[{Van~Doorsselaere {et~al.}(2016)Van~Doorsselaere, Antolin, Yuan,
  Reznikova, \& Magyar}]{10.3389/fspas.2016.00004}
Van~Doorsselaere, T., Antolin, P., Yuan, D., Reznikova, V., \& Magyar, N. 2016,
  Frontiers in Astronomy and Space Sciences, 3, 4,
  \dodoi{10.3389/fspas.2016.00004}

\bibitem[{{Van Doorsselaere} {et~al.}(2008){Van Doorsselaere}, {Nakariakov}, \&
  {Verwichte}}]{2008ApJ...676L..73V}
{Van Doorsselaere}, T., {Nakariakov}, V.~M., \& {Verwichte}, E. 2008, \apjl,
  676, L73, \dodoi{10.1086/587029}

\bibitem[{{Verth} {et~al.}(2010){Verth}, {Terradas}, \&
  {Goossens}}]{2010ApJ...718L.102V}
{Verth}, G., {Terradas}, J., \& {Goossens}, M. 2010, \apjl, 718, L102,
  \dodoi{10.1088/2041-8205/718/2/L102}

\bibitem[{{Walsh} \& {Ireland}(2003)}]{2003A&ARv..12....1W}
{Walsh}, R.~W., \& {Ireland}, J. 2003, \aapr, 12, 1,
  \dodoi{10.1007/s00159-003-0021-9}

\bibitem[{{Weberg} {et~al.}(2018){Weberg}, {Morton}, \&
  {McLaughlin}}]{WEBetal2018}
{Weberg}, M.~J., {Morton}, R.~J., \& {McLaughlin}, J.~A. 2018, \apj, 852, 57,
  \dodoi{10.3847/1538-4357/aa9e4a}

\bibitem[{{Withbroe} \& {Noyes}(1977)}]{1977ARA&A..15..363W}
{Withbroe}, G.~L., \& {Noyes}, R.~W. 1977, \araa, 15, 363,
  \dodoi{10.1146/annurev.aa.15.090177.002051}

\bibitem[{{Young} {et~al.}(2003){Young}, {Del Zanna}, {Landi}, {Dere}, {Mason},
  \& {Landini}}]{2003ApJS..144..135Y}
{Young}, P.~R., {Del Zanna}, G., {Landi}, E., {et~al.} 2003, \apjs, 144, 135,
  \dodoi{10.1086/344365}

\end{thebibliography}

\end{document}